\PassOptionsToPackage{unicode}{hyperref}
\PassOptionsToPackage{hyphens}{url}
\documentclass[
]{article}
\usepackage{xcolor}
\usepackage[margin=1in]{geometry}
\usepackage{amsmath,amssymb}
\usepackage{iftex}
\ifPDFTeX
  \usepackage[T1]{fontenc}
  \usepackage[utf8]{inputenc}
  \usepackage{textcomp} 
\else 
  \usepackage{unicode-math} 
  \defaultfontfeatures{Scale=MatchLowercase}
  \defaultfontfeatures[\rmfamily]{Ligatures=TeX,Scale=1}
\fi
\usepackage{lmodern}
\ifPDFTeX\else
\fi
\IfFileExists{upquote.sty}{\usepackage{upquote}}{}
\IfFileExists{microtype.sty}{
  \usepackage[]{microtype}
  \UseMicrotypeSet[protrusion]{basicmath} 
}{}
\makeatletter
\@ifundefined{KOMAClassName}{
  \IfFileExists{parskip.sty}{%
    \usepackage{parskip}
  }{
    \setlength{\parindent}{0pt}
    \setlength{\parskip}{6pt plus 2pt minus 1pt}}
}{
  \KOMAoptions{parskip=half}}
\makeatother
\usepackage{color}
\usepackage{fancyvrb}

\DefineVerbatimEnvironment{Highlighting}{Verbatim}{commandchars=\\\{\}}
\newenvironment{Shaded}{}{}

\newcommand{\AttributeTok}[1]{\textcolor[rgb]{0.49,0.56,0.16}{#1}}

\newcommand{\BuiltInTok}[1]{\textcolor[rgb]{0.00,0.50,0.00}{#1}}

\newcommand{\CommentTok}[1]{\textcolor[rgb]{0.38,0.63,0.69}{\textit{#1}}}

\newcommand{\ExtensionTok}[1]{#1}

\newcommand{\FunctionTok}[1]{\textcolor[rgb]{0.02,0.16,0.49}{#1}}

\newcommand{\KeywordTok}[1]{\textcolor[rgb]{0.00,0.44,0.13}{\textbf{#1}}}
\newcommand{\NormalTok}[1]{#1}
\newcommand{\OperatorTok}[1]{\textcolor[rgb]{0.40,0.40,0.40}{#1}}

\newcommand{\StringTok}[1]{\textcolor[rgb]{0.25,0.44,0.63}{#1}}
\newcommand{\VariableTok}[1]{\textcolor[rgb]{0.10,0.09,0.49}{#1}}

\usepackage{longtable,booktabs,array}
\usepackage{calc} 
\usepackage{etoolbox}
\makeatletter
\patchcmd\longtable{\par}{\if@noskipsec\mbox{}\fi\par}{}{}
\makeatother
\IfFileExists{footnotehyper.sty}{\usepackage{footnotehyper}}{\usepackage{footnote}}
\makesavenoteenv{longtable}
\providecommand{\tightlist}{%
  \setlength{\itemsep}{0pt}\setlength{\parskip}{0pt}}
\usepackage{caption}
\usepackage{seqsplit}
\usepackage{placeins}

\usepackage{float}

\usepackage{wrapfig}

\usepackage{graphicx}
\usepackage[export]{adjustbox}
\newcommand{\cnxFigMaxW}{\linewidth}
\providecommand{\cnxFigMaxH}{0.42\textheight}
\AtBeginDocument{%
  \let\cnxOrigIncludegraphics\includegraphics
  \renewcommand{\includegraphics}[2][]{%
    \cnxOrigIncludegraphics[#1,max width=\cnxFigMaxW,max height=\cnxFigMaxH,keepaspectratio]{#2}%
  }%
}

\usepackage{needspace}
\usepackage{etoolbox}
\pretocmd{\section}{\Needspace*{4\baselineskip}}{}{}
\pretocmd{\subsection}{\Needspace*{3\baselineskip}}{}{}

\let\origtexttt\texttt
\renewcommand{\texttt}[1]{\origtexttt{\seqsplit{#1}}}

\usepackage{fancyhdr}
\providecommand{\manuShortTitle}{}
\providecommand{\manuHeaderLeft}{}
\fancypagestyle{plain}{%
  \fancyhf{}%
}
\usepackage{newunicodechar}
\newunicodechar{≤}{\ensuremath{\leq}}
\newunicodechar{≥}{\ensuremath{\geq}}
\newunicodechar{≈}{\ensuremath{\approx}}
\newunicodechar{≠}{\ensuremath{\neq}}
\newunicodechar{≡}{\ensuremath{\equiv}}
\newunicodechar{∝}{\ensuremath{\propto}}
\newunicodechar{∼}{\ensuremath{\sim}}
\newunicodechar{≪}{\ensuremath{\ll}}
\newunicodechar{≫}{\ensuremath{\gg}}
\newunicodechar{×}{\ensuremath{\times}}
\newunicodechar{÷}{\ensuremath{\div}}
\newunicodechar{±}{\ensuremath{\pm}}
\newunicodechar{∓}{\ensuremath{\mp}}
\newunicodechar{−}{\ensuremath{-}}
\newunicodechar{∧}{\ensuremath{\wedge}}
\newunicodechar{∨}{\ensuremath{\vee}}
\newunicodechar{¬}{\ensuremath{\neg}}
\newunicodechar{∩}{\ensuremath{\cap}}
\newunicodechar{∪}{\ensuremath{\cup}}
\newunicodechar{∈}{\ensuremath{\in}}
\newunicodechar{∉}{\ensuremath{\notin}}
\newunicodechar{⊂}{\ensuremath{\subset}}
\newunicodechar{⊆}{\ensuremath{\subseteq}}
\newunicodechar{⊇}{\ensuremath{\supseteq}}
\newunicodechar{≅}{\ensuremath{\cong}}
\newunicodechar{□}{\ensuremath{\square}}
\newunicodechar{∀}{\ensuremath{\forall}}
\newunicodechar{∃}{\ensuremath{\exists}}
\newunicodechar{∇}{\ensuremath{\nabla}}
\newunicodechar{∂}{\ensuremath{\partial}}
\newunicodechar{∑}{\ensuremath{\sum}}
\newunicodechar{∏}{\ensuremath{\prod}}
\newunicodechar{∫}{\ensuremath{\int}}
\newunicodechar{√}{\ensuremath{\surd}}
\newunicodechar{∞}{\ensuremath{\infty}}
\newunicodechar{→}{\ensuremath{\rightarrow}}
\newunicodechar{←}{\ensuremath{\leftarrow}}
\newunicodechar{↔}{\ensuremath{\leftrightarrow}}
\newunicodechar{⇒}{\ensuremath{\Rightarrow}}
\newunicodechar{⇐}{\ensuremath{\Leftarrow}}
\newunicodechar{⇔}{\ensuremath{\Leftrightarrow}}
\newunicodechar{↑}{\ensuremath{\uparrow}}
\newunicodechar{↓}{\ensuremath{\downarrow}}
\newunicodechar{α}{\ensuremath{\alpha}}
\newunicodechar{β}{\ensuremath{\beta}}
\newunicodechar{γ}{\ensuremath{\gamma}}
\newunicodechar{δ}{\ensuremath{\delta}}
\newunicodechar{ε}{\ensuremath{\varepsilon}}
\newunicodechar{ζ}{\ensuremath{\zeta}}
\newunicodechar{η}{\ensuremath{\eta}}
\newunicodechar{θ}{\ensuremath{\theta}}
\newunicodechar{κ}{\ensuremath{\kappa}}
\newunicodechar{λ}{\ensuremath{\lambda}}
\newunicodechar{μ}{\ensuremath{\mu}}
\newunicodechar{ν}{\ensuremath{\nu}}
\newunicodechar{ξ}{\ensuremath{\xi}}
\newunicodechar{π}{\ensuremath{\pi}}
\newunicodechar{ρ}{\ensuremath{\rho}}
\newunicodechar{σ}{\ensuremath{\sigma}}
\newunicodechar{τ}{\ensuremath{\tau}}
\newunicodechar{φ}{\ensuremath{\varphi}}
\newunicodechar{χ}{\ensuremath{\chi}}
\newunicodechar{ψ}{\ensuremath{\psi}}
\newunicodechar{ω}{\ensuremath{\omega}}
\newunicodechar{Γ}{\ensuremath{\Gamma}}
\newunicodechar{Δ}{\ensuremath{\Delta}}
\newunicodechar{Θ}{\ensuremath{\Theta}}
\newunicodechar{Λ}{\ensuremath{\Lambda}}
\newunicodechar{Ξ}{\ensuremath{\Xi}}
\newunicodechar{Π}{\ensuremath{\Pi}}
\newunicodechar{Σ}{\ensuremath{\Sigma}}
\newunicodechar{Φ}{\ensuremath{\Phi}}
\newunicodechar{Ψ}{\ensuremath{\Psi}}
\newunicodechar{Ω}{\ensuremath{\Omega}}
\usepackage{newtxtext,newtxmath}

\makeatletter
\newsavebox\pandoc@box
\newcommand*\pandocbounded[1]{%
  \sbox\pandoc@box{#1}%
  \Gscale@div\@tempa{\textheight}{\dimexpr\ht\pandoc@box+\dp\pandoc@box\relax}%
  \Gscale@div\@tempb{\linewidth}{\wd\pandoc@box}%
  \ifdim\@tempb\p@<\@tempa\p@\let\@tempa\@tempb\fi%
  \ifdim\@tempa\p@<\p@\scalebox{\@tempa}{\usebox\pandoc@box}%
  \else\usebox{\pandoc@box}%
  \fi%
}
\makeatother
\usepackage{bookmark}
\IfFileExists{xurl.sty}{\usepackage{xurl}}{} 
\hypersetup{
  pdftitle={CANONIC: Governance Is Compilation},
  pdfauthor={Dexter Hadley, MD/PhD Founder and Chair, CANONIC Foundation · Director of AI, American Board of Precision Medicine founder@canonic.org},
  hidelinks,
  pdfcreator={LaTeX via pandoc}}

\title{CANONIC: Governance Is Compilation}
\usepackage{etoolbox}
\makeatletter
\providecommand{\subtitle}[1]{
  \apptocmd{\@title}{\par {\large #1 \par}}{}{}
}
\makeatother
\subtitle{Constitutional AI Governance --- compiling digital artifacts
into an evidence ledger at scale}
\author{Dexter Hadley, MD/PhD\\
Founder and Chair, CANONIC Foundation · Director of AI, American Board
of Precision Medicine\\
\href{mailto:founder@canonic.org}{\nolinkurl{founder@canonic.org}}}
\date{June 2026}

\begin{document}
\maketitle

\def\manuShortTitle{Constitutional AI Governance}
\def\manuHeaderLeft{Preprint · Hadley 2026}

\section*{Abstract}\label{abstract}
\addcontentsline{toc}{section}{Abstract}

We present CANONIC: governed intelligence that compiles digital
artifacts into an evidence ledger at scale. Large language models
generate prose faster than anyone can check it, the failure Oxford
Languages named \emph{slop}, its 2025 Word of the Year. CANONIC governs
whether content may enter a corpus the way a compiler decides whether a
program is well-formed: mechanically, by a grammar, at the boundary of
admission. Governance reduces to three axioms (Triad, Inheritance,
Introspection) that map one-to-one onto compiler theory's syntax,
scope-resolution, and type-system layers, and admission is a decidable,
linear-time check. We then ask, with a pre-registered cross-provider
benchmark across four regimes, whether structural admission keeps slop
out. It does not: no prose-reading gate reliably separates reliable from
unreliable content. Slop is not a property an algorithm computes. It is
a verdict of domain expertise. So a governance layer does not decide
slop; it keeps the record auditable --- every claim anchored to a
definition, a commit, and an evidence window, reproducible and checkable
end to end.

\subsection{1 Introduction}\label{introduction}

Large language models now generate prose at rates that outpace human
verification capacity, producing a class of artifact that
\href{https://languages.oup.com/word-of-the-year/2025/}{Oxford
Languages} {[}1{]} formalized in December 2025 by elevating the noun
\emph{slop} to its 2025 Word of the Year. The underlying mechanism is
not anomalous behavior but the expected consequence of next-token
prediction over heterogeneous corpora, the objective Vaswani set out in
2017 {[}2{]} and Brown scaled in 2020 {[}3{]}: outputs are optimized for
plausibility rather than truth, and the two objectives diverge whenever
training data is sparse, contested, or absent {[}4{]}. When training
distributions contain the relevant facts, generations tend toward
correctness; when they do not, generations tend toward stylistic
competence detached from referent. The asymmetry has been documented
across summarization, citation generation, and clinical question
answering {[}5{]}.

The downstream cost of this asymmetry scales with the stakes of the
deploying domain. A misgenerated student essay is recoverable at the
level of a single grade, whereas a misgenerated research summary
consumed by a clinician can propagate into a treatment decision, and
policy analyses assembled from unverified machine output can affect
populations {[}6{]}. The prevailing reader model, in which trust is
extended to the author, to the model, to the reviewers, and to the
publishing institution in sequence, was calibrated for a regime of
scarce content and tractable verification, and neither condition is
preserved under contemporary generation rates {[}7{]}. Production now
exceeds validation by orders of magnitude, and the bottleneck of
scholarly communication has inverted from synthesis to audit.

This paper uses \emph{slop} in two senses, and the gap between them is
its organizing finding. The \emph{structural} sense is mechanical:
content never anchored to a referent, as opposed to merely incorrect.
Misinformation admits correction because a truthful referent exists;
structurally unanchored content admits none, because it was never
anchored at all. The \emph{operational} sense, which the evaluation in
§5 makes load-bearing, is the colloquial one: content an expert reader
judges ungrounded or false. Structural anchoring is achievable, and we
measure it, but it does not capture the operational sense, because
fluent fabrication and fabricated data are perfectly anchorable. That
gap is the heart of the paper. Whether content is slop in the
operational sense is a verdict of domain expertise --- a matter of
taste, not a property any algorithm computes. What a machine can do
instead is give that expertise something to converge on: a
\textbf{source of truth} in which every claim resolves to a defined
term, a committed piece of evidence, and a declared window. That source
of truth is the \textbf{evidence ledger} --- an append-only,
git-anchored record of work done, distributed in that anyone can clone
and check it --- and the framework's job is to keep the ledger, not to
render the verdict. Existing defenses instead act only after the
artifact has been produced, and each fails for a structural reason
(Table 1):

\needspace{0.26\textheight}

\textbf{Table 1 \textbar{} Why post-hoc defenses against AI slop fail.}
Each common defense applied after generation, and the structural reason
it cannot keep slop out of the corpus.

{\def\LTcaptype{none} 
\begin{longtable}[]{@{}
  >{\raggedright\arraybackslash}p{(\linewidth - 4\tabcolsep) * \real{0.2468}}
  >{\raggedright\arraybackslash}p{(\linewidth - 4\tabcolsep) * \real{0.2597}}
  >{\raggedright\arraybackslash}p{(\linewidth - 4\tabcolsep) * \real{0.4935}}@{}}
\toprule\noalign{}
\begin{minipage}[b]{\linewidth}\raggedright
Defense
\end{minipage} & \begin{minipage}[b]{\linewidth}\raggedright
Mechanism
\end{minipage} & \begin{minipage}[b]{\linewidth}\raggedright
Why It Fails
\end{minipage} \\
\midrule\noalign{}
\endhead
\bottomrule\noalign{}
\endlastfoot
Detection tools & ``Is this AI-written?'' & Wrong axis --- flags
authorship, not provenance; legitimate work is AI-written too \\
Disclosure policies & Author attestation & ``I used AI responsibly''
proves nothing \\
Human review & Manual inspection & Catches slop but doesn't prevent
it \\
Style guidelines & Surface formatting & Cosmetic; doesn't address
evidence \\
\end{longtable}
}

Each row in the table operates post-hoc on artifacts that already exist,
and none of the listed mechanisms constrains what is admitted to the
corpus.

Each row also asks a variant of the \emph{wrong question}. Detection
asks whether content is AI-written; disclosure asks whether the author
admits it; review and style ask whether it reads well. None of these
survives a regime in which legitimate scholarship is itself
machine-assisted: \textbf{authorship no longer separates sound content
from slop}, because the best work and the worst are now written the same
way. The question that does survive is narrower and structural: \emph{is
every claim anchored to a source of truth?} CANONIC gates on
\textbf{provenance, not authorship}. It never asks whether a passage was
written by a model; it asks whether each claim the passage makes
resolves to a defined term, a committed piece of evidence, and a
declared window --- entries in the same append-only ledger §3 builds, a
source of truth the reader can clone and check. Anchoring is not truth:
a fabricated claim can be perfectly anchored, as §5 shows. But it is the
part a machine can guarantee, and it leaves domain expertise a checkable
record to converge on rather than unverifiable prose. Proving that
anchoring, mechanically and before admission, is the whole of what the
framework guarantees, and §5 shows it is the only axis that holds once
content can no longer be sorted by who or what produced it.

Relocating the verification step from the reader to a gate at the point
of admission has partial precedents that the present work generalizes.
Policy-as-code engines such as Open Policy Agent and HashiCorp Sentinel
admit or reject infrastructure changes against declarative rules
{[}8{]}; provenance standards such as C2PA bind cryptographic assertions
to media at the moment of capture {[}9{]}; model and data cards {[}10{]}
and study preregistration {[}11{]} commit claims to a fixed record
before outcomes are known. Each gates a single slice, a configuration
change, a media file, a disclosure, an experimental protocol. CANONIC
differs in governing the \emph{content artifact itself} against a formal
grammar, at the boundary of the published corpus. One precedent deserves
direct acknowledgment, because it shares this paper's adjective:
Constitutional AI {[}12{]}, the alignment method introduced by Bai and
colleagues at Anthropic, the company Dario Amodei co-founded and leads.
The names are close. The mechanisms are not. Anthropic's constitution is
a \emph{training signal}. A model critiques and revises its own outputs
against a set of natural-language principles, and those AI-generated
preferences are distilled into the weights by reinforcement learning
from AI feedback. That constitution is model-scoped, absorbed into
parameters, and leaves no audit trail when it changes. CANONIC leaves
the model unconstrained. It governs whether \emph{content} is admissible
to a corpus, through formal, versioned constraints checked at the
compiler boundary, and every change to its constitution is a git diff
anyone can inspect. Anthropic's constitution trains the model. CANONIC's
governs the institution {[}13{]}. The locus of control is the corpus
boundary, not the decoder. We position CANONIC against this and the
other related systems in §6.

\textbf{Our contributions are:}

\begin{enumerate}
\def\labelenumi{\arabic{enumi}.}
\tightlist
\item
  \textbf{A reframing.} We recast \emph{governance as compilation}:
  admission as a decidable decision about well-formedness made at the
  corpus boundary before publication, separate in kind from a judgment
  about truth or quality made afterward.
\item
  \textbf{Three axioms and a compiler correspondence.} We give a
  three-axiom governance basis (Triad, Inheritance, Introspection) and
  show it maps one-to-one onto the three pillars of compiler theory,
  syntax, scope resolution, type systems (§4, Theorems 1 through 3),
  with the axioms shown mutually independent (Theorem 4) and
  individually load-bearing (Theorem 5).
\item
  \textbf{A decidable admission procedure.} We specify
  \texttt{validate(scope)} (Algorithm 1) and prove scope validity
  decidable in linear time (Theorem 6), so admission is a mechanical
  pass or fail rather than a reviewer's judgment.
\item
  \textbf{A git-anchored construction.} We report a one-month build of
  ten repositories and twenty governed scopes in which every count
  resolves to a public commit, and we mark plainly what the gates do not
  do: they bound the unaccountable, not the false (§3, §7).
\item
  \textbf{An adversarial evaluation of what the gates can and cannot
  do.} We build a pre-registered, cross-provider benchmark and show,
  across non-adversarial synthetic, adversarial, novel-domain, and
  real-world content, that structural admission, retrieval grounding,
  and semantic judgment are each statistically independent of, or
  defeatable on, truth, so the framework's guarantee is
  \emph{accountability}, not filtering (§5). This converts the boundary
  of §3 from an assertion into a measurement, and supplies a map of
  where automated truth-filtering fails.
\end{enumerate}

\subsection{2 Background: Governance as
Compilation}\label{background-governance-as-compilation}

Compiler theory addressed a structurally analogous validation problem
more than half a century ago, when programs that appeared to execute
correctly were observed to corrupt data silently, fail under
unanticipated inputs, or terminate without diagnostic, and ad-hoc
runtime checking was found to scale poorly against the combinatorial
space of programmer error. The response, formalized through
\href{https://doi.org/10.1016/S0019-9958(59)90362-6}{Chomsky's hierarchy
of formal grammars} {[}14{]} and operationalized by
\href{https://doi.org/10.1016/S0019-9958(65)90426-2}{Knuth and the
LR-parsing tradition that mechanized it} {[}15{]}, reframed program
validity as a property to be decided at the language level rather than
discovered during execution. A program omitting a required token,
referencing an undefined identifier, or violating type or scope
constraints is rejected at compile time and never reaches the runtime;
the canonical treatment by
\href{https://suif.stanford.edu/dragonbook/}{Aho, Lam, Sethi, and
Ullman} {[}16{]} establishes lexical analysis, parsing, semantic
analysis, and intermediate-representation checking as the standard
pipeline through which structural validity is enforced before any
execution semantics apply. The same construction is available to
governance. If the structural requirements for valid content are
specified as a formal grammar over typed artifacts, and if validators
implementing that grammar are interposed before any artifact is admitted
to the corpus, then structurally deficient content does not enter a
degraded state in which it must later be detected, contested, or
retracted; it is \emph{malformed} in the precise compiler-theoretic
sense, and the rejection is mechanical rather than evaluative.
Governance is, under this reframing, compilation over a content language
whose well-formedness conditions encode the institutional requirements
that post-hoc review has historically been asked to enforce. CANONIC is
the compiler-theoretic instantiation of that reframing (first set out in
an early note on the compiler insight {[}17{]}), and its grammar reduces
to three axioms developed in §3.

\subsection{3 Method: The Three Axioms}\label{method-the-three-axioms}

CANONIC's framework derives from three rules that we refer to throughout
this paper as the \textbf{Triad}, \textbf{Inheritance}, and
\textbf{Introspection} axioms, and which together constitute the
constitutional foundation on which every other property of the system
rests. We call them axioms rather than rules or principles because they
cannot be derived from anything more primitive within the framework
itself; they must be asserted as foundation, much as a dictionary such
as Oxford Languages cannot derive the meaning of a word from the letters
that compose it, and must instead anchor meaning in an external
definitional act. Figure 1 shows the four-element closure model that
these three axioms together induce.

\begin{figure}[!ht]
\centering

\pandocbounded{\includegraphics[keepaspectratio,alt={}]{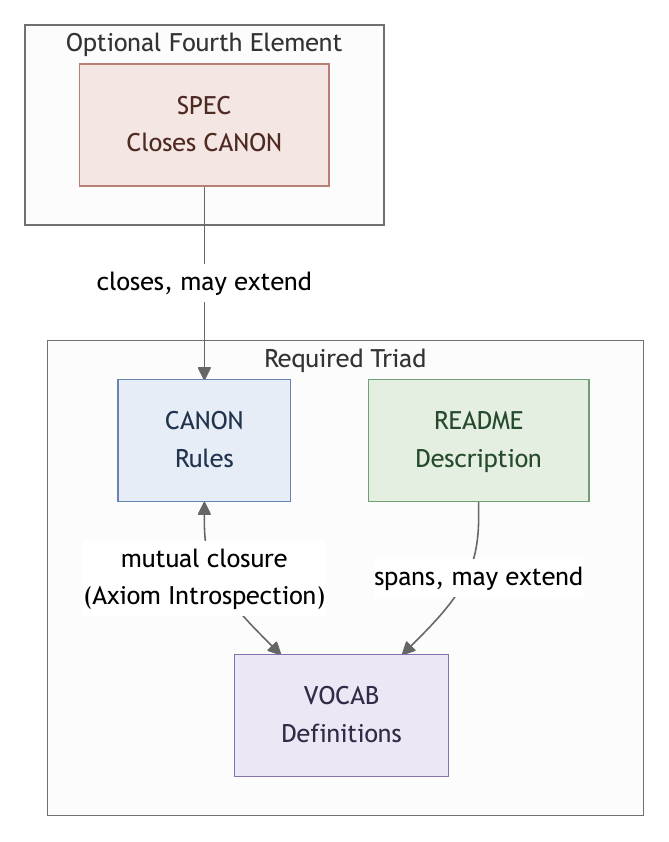}}

\textbf{Figure 1: Triad and SPEC closure model for a governed scope.}
Every scope contains the three required files of the \emph{Triad}:
\texttt{CANON.md} (rules), \texttt{VOCAB.md} (definitions), and
\texttt{README.md} (description). An optional fourth file, \emph{SPEC}
(named \texttt{CANONIC.md} at the project root, or \texttt{PAPER.md} for
this manuscript), closes \texttt{CANON.md} and may extend it with
generation details. Arrows derive from the introspection axiom:
\texttt{CANON.md} and \texttt{VOCAB.md} are mutually closed,
\texttt{README.md} spans \texttt{VOCAB.md} and may add terms, and
\emph{SPEC} closes \texttt{CANON.md} and may add generation rules. A
scope missing any \emph{Triad} element is structurally invalid and
cannot compile.

\end{figure}

\textbf{Axiom 0, Triad.} A governed unit MUST contain three files:
\texttt{CANON.md} (governance), \texttt{VOCAB.md} (semantics), and
\texttt{README.md} (description).

\textbf{Axiom 1, Inheritance.} Every \texttt{CANON.md} MUST declare its
parent scope. Inheritance chains MUST terminate at root. Inherited rules
are final.

\textbf{Axiom 2, Introspection.} \texttt{VOCAB.md} MUST define every
concept used in \texttt{CANON.md} and in \texttt{VOCAB.md} itself.

These three axioms are put forward as a \emph{sufficient} governance
basis for the corpus reported here rather than a proven-complete one.
Two of the three properties we claim for them are discharged in Appendix
B by explicit counterexample: they are \emph{independent} (Theorem 4,
none is derivable from the others) and \emph{minimal} (Theorem 5,
removing any one breaks a closure property developed below). The third
property, \emph{exhaustiveness}, that every further governance rule
reduces to a consequence of the three, is an empirical observation over
our roughly one-month window, not a theorem; we state it as a design
hypothesis and flag in §7 that a rule outside the three would falsify
it.

We unpack each axiom in turn.

\subsubsection{The Triad axiom}\label{the-triad-axiom}

A governed unit is called a scope, and a scope is simply a directory in
a filesystem. The triad is the minimal set of files that make governance
possible (Table 2):

\needspace{0.26\textheight}

\textbf{Table 2 \textbar{} The three required files of a governed scope
(the Triad).} CANON, VOCAB, and README, and the role each plays.

{\def\LTcaptype{none} 
\begin{longtable}[]{@{}lll@{}}
\toprule\noalign{}
File & Role & Contains \\
\midrule\noalign{}
\endhead
\bottomrule\noalign{}
\endlastfoot
\textbf{CANON.md} & Governance & Axioms: what MUST be true \\
\textbf{VOCAB.md} & Semantics & Definitions: what terms mean \\
\textbf{README.md} & Description & Documentation: what this is \\
\end{longtable}
}

Three files are required because governance requires a separation of
concerns. The canon file is normative and prescribes the rules; the
readme file is descriptive and explains the purpose of the scope; the
vocabulary file bridges the two by ensuring that the rules use defined
terms. Collapsing these roles into a single document invites predictable
failure modes, in which documentation accidentally governs, governance
accidentally describes, or the same term carries different meanings in
different parts of the tree.

The triad therefore enforces separation by construction. A scope that is
missing any of the three files is not merely incomplete or deprecated;
it is structurally malformed, and the compiler refuses to build it.

\subsubsection{The Inheritance axiom}\label{the-inheritance-axiom}

Scopes exist in hierarchies. A scope at \texttt{/services/writing/}
inherits from \texttt{/services/}, which in turn inherits from
\texttt{/} at the root. The inheritance chain defines the flow of
authority, and child scopes cannot override the axioms of their parents.

The finality of inherited axioms is a constitutional necessity rather
than a design preference, and it is worth pausing to consider why a more
permissive arrangement would be incompatible with the governance
properties we want the framework to guarantee. If child scopes were free
to override the axioms of their parents, authority would fragment across
the tree, and different branches of the hierarchy could establish
mutually contradictory rules without any mechanism for detecting or
reconciling the contradiction. The resulting structure would not in any
meaningful sense be a governed system, since inheritance combined with
arbitrary override reduces to anarchy that has merely accumulated
additional ceremony.

Inherited axioms are therefore final within the scope of any descendant,
and this finality is what allows the framework to make stable guarantees
about properties such as evidence-traceability and vocabulary-closure
across the entire governed tree. A child scope retains the freedom to
\emph{extend} governance by introducing new axioms of its own that apply
locally to its own descendants, but it cannot modify or relax the axioms
that it inherits from any ancestor in the chain that runs back to the
root. The constitution declared at the root of the tree therefore binds
every descendant scope without exception, which is the precise sense in
which we mean that the framework is constitutional rather than merely
conventional.

\subsubsection{The Introspection axiom}\label{the-introspection-axiom}

The vocabulary must define itself. If a canon file uses the term
\emph{scope}, the vocabulary file at the same scope must contain a
definition of \emph{scope}. If that definition itself uses the term
\emph{axiom}, the vocabulary file must also define \emph{axiom}. The
closure is reflexive: the vocabulary is required to cover not only the
canon it serves but also the language in which it is itself written.

The effect is to eliminate \emph{undefined} terms, jargon drift, and
authority borrowed from concepts that mean different things to different
readers: every term used in governed text must resolve to a definition
at the appropriate scope, and the validator rejects any reference whose
target is missing. Introspection therefore makes content accountable to
a stated vocabulary. It does not, however, make slop
\emph{inadmissible}; a producer can satisfy closure vacuously with
circular definitions, since the gate checks that a term is
\emph{defined}, not that the definition is \emph{informative}. We name
this and the other vacuous-satisfaction attacks below, and in §5 we
measure their effect: structural admission turns out to carry no
information about whether an admitted claim is true.

\subsubsection{The validation gates}\label{the-validation-gates}

Claims move through three gates. Failure at any gate renders the content
invalid. Figure 2 traces the pipeline from a candidate claim through
three structural checks to either ACCEPT or REJECT.

\begin{figure}[!ht]
\centering

\pandocbounded{\includegraphics[keepaspectratio,alt={}]{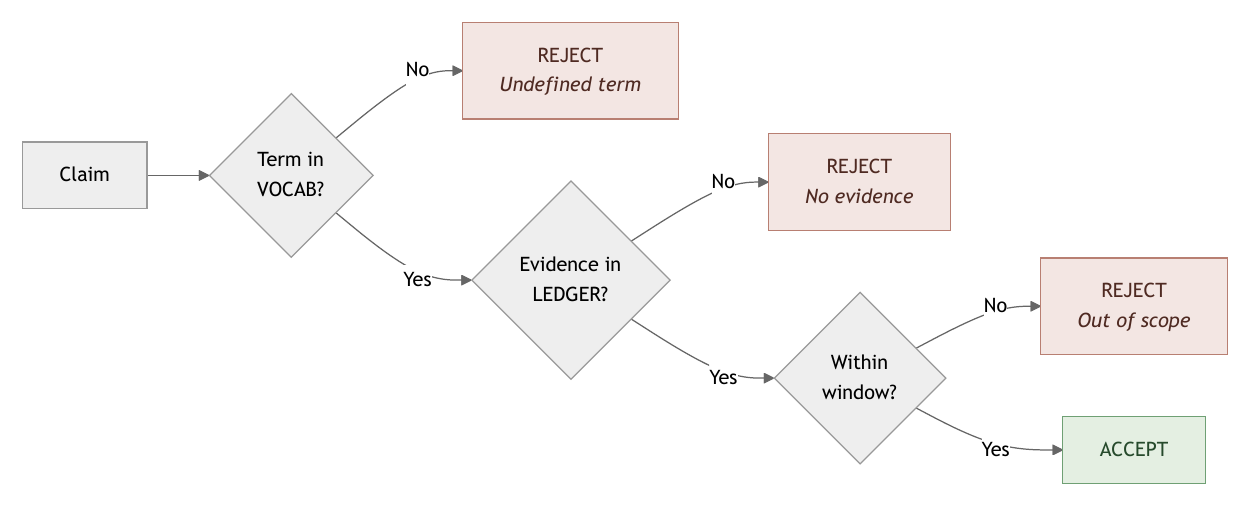}}

\textbf{Figure 2: The three-gate validation pipeline.} Each claim is
checked at Vocabulary (whether the term is defined), Evidence (whether a
ledger entry exists), and Scope (whether the evidence falls within the
declared window). Failure at any gate produces a REJECT with a
structural reason, and only claims that pass all three reach ACCEPT.

\end{figure}

\textbf{Gate 1: Vocabulary.} Does the claim use terms defined in the
scope's \texttt{VOCAB.md}? Undefined jargon fails at this gate, and the
model cannot introduce terminology that humans have not sanctioned.

\textbf{Gate 2: Evidence.} Does the claim cite something in the ledger?
A ledger is an append-only record, implemented here as a
\href{https://git-scm.com/}{git repository}. Assertions without commits
carry no evidence, and they fail.

\textbf{Gate 3: Scope.} Is the evidence within the declared window?
Every paper declares its evidence bounds, and claims about evidence
outside that window are inadmissible.

The gates are structural rather than evaluative.
\href{https://github.com/canonic-machine/VALIDATORS}{Validators} do not
assess quality. They check presence. Either a term is defined or it is
not, and either evidence exists or it does not. Binary validity
eliminates judgment calls.

Structural admission is the load-bearing mechanism of the framework, and
it is worth distinguishing carefully from the post-hoc quality filters
that downstream consumers of AI-generated content typically rely upon.
The structural gate is not a quality filter applied after the fact;
malformed content is rejected at the gate itself, so \emph{unanchored}
material, a claim with no defined term, no recorded evidence, no
declared provenance, never enters the corpus and never needs to be
filtered out of it. The principle, and its limit, are exactly a
compiler's: a compiler does not evaluate whether the code it receives is
\emph{correct}, only whether it is structurally valid against the
language specification, and CANONIC inherits both that guarantee and
that boundary. A governed artifact compiles when every claim it makes is
anchored, to a definition, to a commit, to the declared window, and
admission is a binary compilation outcome that either succeeds or fails.
What admission is \emph{not} is a warrant of truth. A false statement
that is nonetheless anchored to a mistaken or fabricated source will
compile, exactly as a well-typed program can still compute the wrong
answer. The gate bounds what can be \emph{unaccountable}, not what can
be \emph{wrong}; we make this boundary precise, and adversarial, in the
next subsection and again in §7.

\subsubsection{Adversarial admission}\label{adversarial-admission}

The gates check for presence, and a presence check rewards whoever
supplies presence. The property that makes admission decidable also
bounds what it can guarantee: an adversary who knows the grammar can
satisfy it vacuously. Three attacks are worth naming because each
defeats one gate without supplying the content the gate is meant to
stand in for. \emph{Vocabulary padding} defeats the introspection gate,
a producer can define every term it uses, however circularly (``X:
guidance pertaining to X''), and close the gate without conveying
meaning, because closure verifies that a term is \emph{defined}, not
that the definition is informative. \emph{Evidence fabrication} defeats
the anchoring gate, the gate verifies that a cited commit exists, not
that the commit records a true observation, so a hallucinated finding
written into a real commit compiles, now bearing a hash and the false
authority that a hash confers. \emph{Window shopping} defeats the
evidence-window constraint, because the window is author-declared, a
producer can choose the interval that flatters the claim. None of these
is hypothetical for the corpus reported here. Its evidence is an n=1,
self-authored artifact validated by validators the authors wrote, on
content the authors had every incentive to make compile (§7); producer,
validator, and evidence are the same party, which is the weakest
possible adversary model. CANONIC raises the cost of
\emph{unaccountable} content, every claim must name a source that
exists, but it does not raise the cost of \emph{accountable falsehood}.
Closing that gap is not a matter of adding more presence checks. It
requires validators that score the \emph{informativeness} of a
definition and the \emph{fidelity} of evidence to the claim it anchors,
together with independent parties to run them. §5 measures exactly this
gap with an independent, cross-provider benchmark and finds it large:
structural admission is statistically independent of truth, and an
adversary optimizing against the semantic check halves its catch rate.
We regard closing that gap, not any figure in the construction census,
as the framework's principal open problem.

\subsubsection{Provenance: how the axioms were
discovered}\label{provenance-how-the-axioms-were-discovered}

The three axioms emerged from practice, not theory. On December 29,
2025, the first governance file (named \texttt{CANNON.md}, later
corrected to \texttt{CANON.md} once the doubled consonant was caught)
was committed at \texttt{07a5834} to the \emph{Dividends \& Deaths}
repository, a book on the economics of precision medicine {[}17{]}. The
conversation that produced it ran in OpenAI's ChatGPT, content-addressed
in the evidence store with its sha256 digest recorded in the Evidence
Ledger (§ Context Anchors). The genesis is therefore anchored twice
over: the commit fixes the canon, and the hash fixes the context that
authored it. It declared three rules for a fast-growing manuscript: law
is separate from narrative; an uppercase filename signals cross-asset
scope; prose is canonical only if it traces to an asset. Drafted for one
book, the rules generalized over the month that followed to govern any
structured content under model authorship. By the window close they had
induced governed \emph{analogs} of roughly fifty years of computing
infrastructure: ten repositories and twenty governed scopes standing in
for an operating system, an immutability layer, a distribution channel,
and a token economy. The claim is deliberately narrow. A scope named for
a token economy declares the rules such an economy must obey. It does
not mint tokens. What the three rules compiled into was structure, not
systems. The genesis commit and each layer's founding commit are
git-anchored in Table 4.

\subsection{4 Formalization}\label{formalization}

The CANONIC parallel to compiler theory is not a loose analogy. It is a
formal correspondence. We state the load-bearing objects here and defer
the proofs to Appendix B. A scope is the unit of governance: a directory
carrying the Triad, a declared parent, and a set of definitions. A scope
is \emph{valid} when its Triad is present, its inheritance chain
terminates at the root, and every term it uses resolves to a definition
along that chain (Definition 4, Appendix B). Admission is the decision
of that predicate, and it is decidable in one linear pass.

\textbf{Algorithm 1} is the admission procedure. Its three phases
realize the three axioms in order: a constant-time check that the Triad
is present (syntax), a bounded walk to the root with cycle detection
(scope resolution), and a finite test that every term resolves to a
definition over the inheritance closure (type system).

\begin{verbatim}
Algorithm 1  validate(scope) — structural admission of a governed scope
Input:  scope S = (P, T, A, V)        # path, triad, axioms, definitions
Output: VALID | INVALID
Cost:   O(n), n = total size of files in the inheritance chain

 1  # Triad (syntax):              O(1)
 2  if not {CANON.md, VOCAB.md, README.md} subset of files(P):
 3      return INVALID
 4  # Inheritance (scope resolution):  O(d), d = chain depth
 5  chain := []
 6  current := S
 7  while current != root:
 8      if current in chain: return INVALID      # cycle never reaches root
 9      chain.append(current); current := parent(current)
10  # Introspection (type system):  O(n)
11  defs  := union(definitions(v) for v in chain)
12  terms := union(terms(c) for c in chain) | union(terms(v) for v in chain)
13  if not terms subset of defs: return INVALID  # a term resolves to no definition
14  return VALID
\end{verbatim}

The check is binary and evaluative of presence, not quality: it admits
anchored content and rejects unanchored content, but it does not assess
whether an anchored claim is true (§3, the adversarial-admission
subsection). The three governance axioms (Triad, Inheritance, and
Introspection) map one-to-one onto the three pillars of classical
compiler design: syntax, scope resolution, and the type system (Table
3).

\needspace{0.26\textheight}

\textbf{Table 3 \textbar{} The three CANONIC axioms mapped one-to-one
onto classical compiler-theory pillars.}

{\def\LTcaptype{none} 
\begin{longtable}[]{@{}
  >{\raggedright\arraybackslash}p{(\linewidth - 4\tabcolsep) * \real{0.2099}}
  >{\raggedright\arraybackslash}p{(\linewidth - 4\tabcolsep) * \real{0.1975}}
  >{\raggedright\arraybackslash}p{(\linewidth - 4\tabcolsep) * \real{0.5926}}@{}}
\toprule\noalign{}
\begin{minipage}[b]{\linewidth}\raggedright
CANONIC Axiom
\end{minipage} & \begin{minipage}[b]{\linewidth}\raggedright
Compiler Concept
\end{minipage} & \begin{minipage}[b]{\linewidth}\raggedright
Function
\end{minipage} \\
\midrule\noalign{}
\endhead
\bottomrule\noalign{}
\endlastfoot
\textbf{Triad} & Syntax & What structures must exist in valid
programs \\
\textbf{Inheritance} & Scope Resolution & Where names resolve; binding
rules \\
\textbf{Introspection} & Type System & What terms must be defined;
semantic constraints \\
\end{longtable}
}

These pillars are textbook material.
\href{https://doi.org/10.1109/TIT.1956.1056813}{Chomsky formalized
syntax in 1956} {[}18{]},
\href{https://doi.org/10.1145/942582.807990}{Johnston gave the contour
model for block-structured scope resolution in 1971} {[}19{]}, and
\href{https://doi.org/10.1016/0022-0000(78)90014-4}{Milner published a
theory of type polymorphism in 1978} {[}20{]}; the unified treatment in
Aho, Lam, Sethi, and Ullman {[}16{]}, the Dragon Book, has been
classroom canon for two decades. The three axioms recover the structure
of these pillars, reached through governance constraints rather than
through programming-language research. A language specification defines
which programs are valid; the CANONIC validators define which governance
structures are valid by analogous machinery, so the correspondence is
structural rather than merely terminological. It is, however, a
correspondence and not an isomorphism, and Appendix B is careful about
the difference: the content ``grammar'' of the triad is a finite
well-formedness checklist rather than a recursive language, and the
type-system analogy shares the \emph{closure} property of the
Hindley--Milner system without its inference. What the correspondence
buys is a precise vocabulary and a decidable admission procedure
(Appendix B.4), not a claim that governance and compilation are the same
object.

This correspondence carries a corollary: governance axioms can generate
computing infrastructure.

The corresponding historical layers in computing infrastructure accreted
over decades: the Unix kernel in 1969, POSIX operating-system standards
in 1988, the iOS App Store distribution model in 2008, and
\href{https://bitcoin.org/bitcoin.pdf}{Bitcoin's immutable ledger} in
2009, each requiring years of development, standardization, and
adoption. CANONIC derived analogous structures on a much shorter
timeline, and the claim is checkable: each layer below is anchored to
the first commit that created it in the public \texttt{canonic-machine}
organization, so the date is the commit's date, not a rounded estimate
(Table 4).

\needspace{0.26\textheight}

\textbf{Table 4 \textbar{} Computing-infrastructure layers anchored to
their founding commits. The genesis --- the original \texttt{CANNON.md},
before the doubled consonant was corrected --- is in the
\texttt{hadleylab/Dividends-and-Deaths} proto-CANONIC repository; the
CANONIC layers follow in the public CANONIC GitHub organizations. Each
date is the first commit in the named repository (verify with
\texttt{git\ log\ -\/-reverse}), making the compressed-timeline claim
reproducible rather than rhetorical.}

{\def\LTcaptype{none} 
\begin{longtable}[]{@{}
  >{\raggedright\arraybackslash}p{(\linewidth - 6\tabcolsep) * \real{0.3191}}
  >{\raggedright\arraybackslash}p{(\linewidth - 6\tabcolsep) * \real{0.1064}}
  >{\raggedright\arraybackslash}p{(\linewidth - 6\tabcolsep) * \real{0.4681}}
  >{\raggedright\arraybackslash}p{(\linewidth - 6\tabcolsep) * \real{0.1064}}@{}}
\toprule\noalign{}
\begin{minipage}[b]{\linewidth}\raggedright
Layer
\end{minipage} & \begin{minipage}[b]{\linewidth}\raggedright
Historical
\end{minipage} & \begin{minipage}[b]{\linewidth}\raggedright
Founding commit (\texttt{org/repo} @ \texttt{sha})
\end{minipage} & \begin{minipage}[b]{\linewidth}\raggedright
Date
\end{minipage} \\
\midrule\noalign{}
\endhead
\bottomrule\noalign{}
\endlastfoot
CANNON (proto-CANONIC genesis) & --- &
\texttt{hadleylab/Dividends-and-Deaths} @ \texttt{07a5834} &
2025-12-29 \\
Kernel (the three axioms) & --- & \texttt{canonic-machine/canonic} @
\texttt{11affab} & 2026-01-05 \\
Operating System & decades & \texttt{canonic-machine/OS} @
\texttt{79bc277} & 2026-01-10 \\
Immutability layer & years & \texttt{canonic-machine/LEDGER} @
\texttt{d578676} & 2026-01-10 \\
Distribution / composition & years & \texttt{canonic-machine/STACK} @
\texttt{f58ad6d} & 2026-01-12 \\
Token economy & years & \texttt{canonic-magic/.canonic} @
\texttt{32fe81f} & 2026-01-30 \\
\end{longtable}
}

The genesis row anchors §3: the first \texttt{CANNON.md} landed in the
\emph{Dividends \& Deaths} book repository on December 29, 2025. Read
top to bottom, the table is a month of computing history in
fast-forward, kernel, operating system, immutability, distribution,
closing with the token economy, whose COIN primitive entered governance
on January 30. Every row resolves to a commit a reader can check. The
date is the founding commit, not a rounded estimate. The token economy
is simply the last era to arrive, the layer that closes the month, not
an exception bolted on after the fact.

The compression is not the result of building five systems in parallel
but of deriving five views from one kernel. The three axioms, applied
recursively, generate the structure each layer requires, so every
additional infrastructure layer is a governed scope inheriting from its
parent rather than an independent engineering effort. What collapses
into three weeks is the governance \emph{skeleton} of decades of
compiler-theoretic and systems work, the shared structure of
declaration, scope, and closure, recovered by composition, not the
implementations themselves.

\subsubsection{The language
specification}\label{the-language-specification}

CANONIC v0.1 includes a formal language specification.
\href{https://github.com/canonic-machine/canonic}{\texttt{LANGUAGE.md}
v0.1} is the formal specification of governed content.

The specification follows conventions from established
programming-language designs, those of Ritchie {[}21{]}, Gosling
{[}22{]}, Bradbury {[}23{]}, and Klabnik {[}24{]}, while introducing
governance-specific constructs.

\subsubsection{The scope-template
grammar}\label{the-scope-template-grammar}

Every governed directory follows a template:

\begin{verbatim}
{SCOPE}/
    CANON.md      -- what MUST be (LAW)
    VOCAB.md      -- what words mean
    README.md     -- what this is
    COVERAGE.md   -- what's missing
    {SCOPE}.md    -- SPECialized SCOPE (STORY)
\end{verbatim}

The filename \texttt{\{SCOPE\}.md} is a template variable; the directory
name becomes the specification filename. For directory \texttt{paper/},
the specification is \texttt{PAPER.md}. For directory
\texttt{validators/}, the specification is \texttt{VALIDATORS.md}.

The naming convention creates a self-referential closure that we
consider load-bearing for the broader claim that the framework is
genuinely constitutional rather than merely procedural. The root
directory of the system is \texttt{canonic/}, and the specification that
governs that root is consequently \texttt{CANONIC.md}, which means that
the framework which governs governance is itself a governed artifact
subject to the same axioms it imposes on its descendants. The language
can therefore describe its own scope and submit to its own verifiers
without invoking any external authority, and the resulting closure is
structural in the sense that it is enforced by the compiler rather than
rhetorical in the sense that it merely claims to hold.

Within each governed scope the two files play complementary roles, since
\texttt{CANON.md} carries the law of that scope while
\texttt{\{SCOPE\}.md} carries the story that makes that law intelligible
to a human reader who arrives without prior context. The
\texttt{CANON.md} file contains axioms, by which we mean the normative
rules that govern the scope and that state what must be true of any
artifact admitted under it, and it deliberately excludes the lifecycle
history, prior rationale, or evolutionary commentary that a reader might
otherwise expect in a single combined document. The companion
\texttt{\{SCOPE\}.md} file closes that gap by providing exactly the
context that makes the axioms reproducible in practice, including the
scope's purpose, its lifecycle phase, the evidence window over which its
claims are valid, and the validation instructions that allow an
independent reader to confirm the claims. The bundle
\texttt{\{CANON.md,\ \{SCOPE\}.md\}} therefore constitutes the minimal
reproducible governance unit of the framework, and neither file is
complete or interpretable without the other.

\subsubsection{The governance loop}\label{the-governance-loop}

Production under CANONIC follows a closed loop. The producing model
generates artifacts, and CANONIC validators check the three axioms
(Triad, Inheritance, Introspection). Any output that would otherwise be
AI slop is rejected before it leaves the gate. The pattern recurs at
every layer: humans declare canon, AI produces artifacts, validators
check against canon, and the ledger records the result. Nothing enters
production without passing validation, and nothing in the ledger lacks a
human-authored canon entry behind it. Figure 3 diagrams this loop and
the one-way arrow that holds it together.

\begin{figure}[!ht]
\centering

\pandocbounded{\includegraphics[keepaspectratio,alt={}]{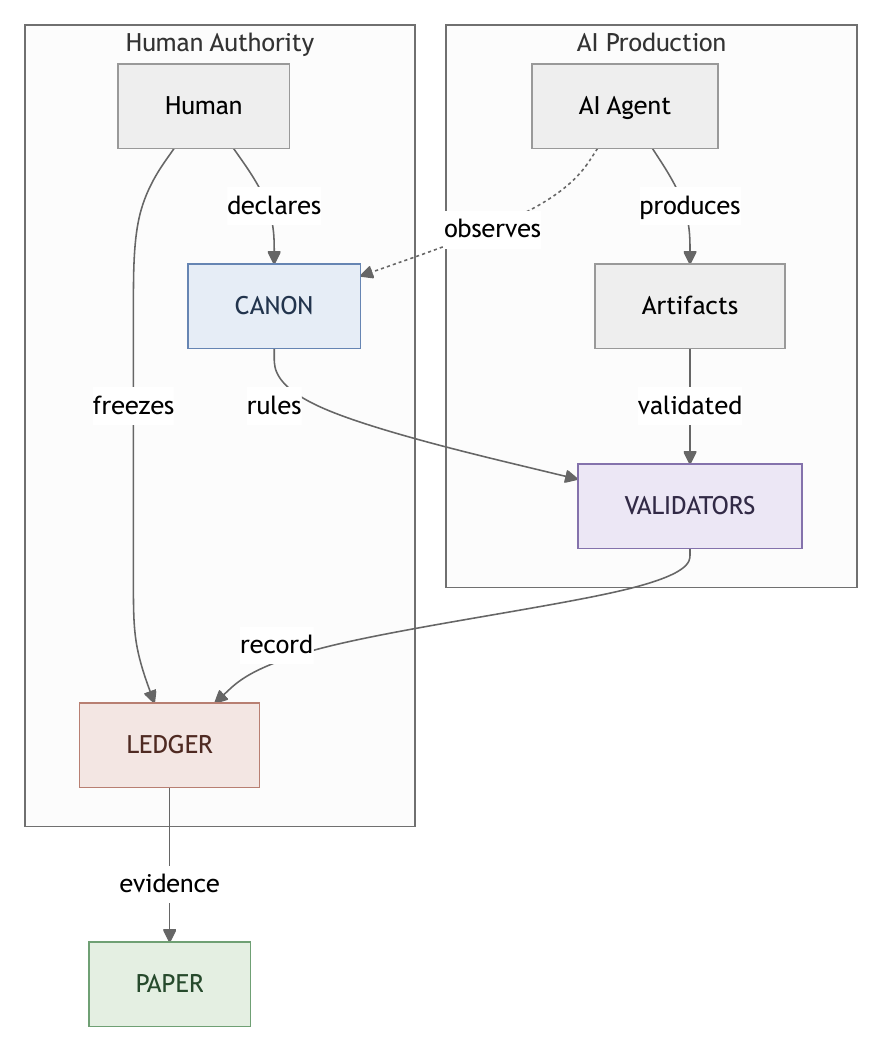}}

\textbf{Figure 3: The AI-First, Human-Governed loop.} Humans declare
CANON and freeze LEDGER. AI produces Artifacts that flow through
VALIDATORS (which read CANON's rules); validated artifacts record into
LEDGER; PAPER cites only LEDGER-resolved evidence. The dotted line, AI
\emph{observes} CANON, marks the architectural asymmetry.

\end{figure}

The dotted line records the architectural asymmetry of the system. AI
observes canon but cannot modify it, and this is not a permission
setting the operator could flip. Governance files live in
human-controlled repositories, and AI sessions read those repositories
while writing only to governed workspaces. Write access to canon is
exclusive to humans, and that exclusivity is constitutional rather than
configurable. We call this arrangement \emph{AI-First, Human-Governed}.
The corollary is that AI can draft, propose, analyze, and critique
without limit, but it cannot canonify. Canonification, the act of
elevating an observed pattern to law, is reserved for human judgment,
and before that act a pattern is a suggestion rather than a rule.

The asymmetry has a temporal edge, and the genesis makes it concrete.
That first conversation ran in OpenAI's ChatGPT, and the framework month
that followed was produced with Anthropic's Claude, the same migration
the wider market made over the same window. The producing agent was
never constant, and it left no binding trace when it turned over. What
persisted instead is the pair of anchors introduced with the genesis
above: the commit that fixes the canon and the content hash that fixes
the context that authored it, the first held in git and the second in
the content-addressed store, both public and both durable. No line of
canon moved when the model changed. Governance, on this reading, is not
a constraint laid over production but the form production's evolution
takes. The agent turns over and the market turns over with it, while the
governed corpus is what survives the turnover, citable to the commit and
to the hash alike.

\subsection{5 Evaluation}\label{evaluation}

We bound every claim to an evidence window of December 29, 2025 through
January 30, 2026, roughly a month from the day after Oxford Languages
named ``AI slop'' Word of the Year. Every assertion below is bounded by
that window, and the full ledger of AI-collaboration sessions inside it
is public, from the ChatGPT genesis through the Claude construction
month.

Claims pertain only to evidence within this window. Bounding the scope
lets readers know exactly what the paper covers and verify coverage
independently.

\subsubsection{Does structural admission filter slop? A controlled
benchmark}\label{does-structural-admission-filter-slop-a-controlled-benchmark}

The construction reported below shows that the framework \emph{validates
itself}: every scope satisfies the three axioms. But a scope passing its
own validators is no more evidence that the gate keeps slop out than a
program compiling is evidence that it is correct; and the construction
is, by §3's own admission, n=1 and self-authored, with producer,
validator, and evidence the same party. We therefore evaluate what the
gates \emph{do} with a separate, pre-registered, cross-provider
benchmark (June 2026), declared here as a later measurement that
modifies no window-close count.

Admission is framed as a binary classifier and scored by its
\textbf{association with content-truth} --- the Matthews correlation φ
between a gate's accept/reject decision and an independent slop/legit
label, where φ = 0 means the decision carries no information about truth
--- rather than by self-reported pass rates. Labels come from a
three-family panel (DeepSeek, OpenAI gpt-4o, Anthropic claude-sonnet)
judging prose alone (Fleiss' κ = 0.91); adversarial content is produced
by a \emph{different} model family than the one that authored CANONIC;
and the structural validators are re-implemented from the published
axioms (Appendix A), so producer, validator, and judge are no longer the
same party. We compare four gates --- structural admission (the three
axioms, \textbf{T}), retrieval-similarity grounding against the governed
corpus (\textbf{B3}), retrieve-then-entail grounding (\textbf{B3e}), and
a frontier semantic judge (\textbf{B2}) --- across four regimes (Table
5).

\needspace{0.26\textheight}

\textbf{Table 5 \textbar{} Gate association with content-truth (Matthews
φ; 0 = independent of truth) across four regimes.} Cross-provider,
pre-registered; 95\% CIs and per-source detail in the released harness.

{\def\LTcaptype{none} 
\begin{longtable}[]{@{}
  >{\raggedright\arraybackslash}p{(\linewidth - 8\tabcolsep) * \real{0.4194}}
  >{\raggedright\arraybackslash}p{(\linewidth - 8\tabcolsep) * \real{0.1183}}
  >{\raggedright\arraybackslash}p{(\linewidth - 8\tabcolsep) * \real{0.1398}}
  >{\raggedright\arraybackslash}p{(\linewidth - 8\tabcolsep) * \real{0.2043}}
  >{\raggedright\arraybackslash}p{(\linewidth - 8\tabcolsep) * \real{0.1183}}@{}}
\toprule\noalign{}
\begin{minipage}[b]{\linewidth}\raggedright
Regime
\end{minipage} & \begin{minipage}[b]{\linewidth}\raggedright
T structure
\end{minipage} & \begin{minipage}[b]{\linewidth}\raggedright
B3 similarity
\end{minipage} & \begin{minipage}[b]{\linewidth}\raggedright
B3e retrieve+entail
\end{minipage} & \begin{minipage}[b]{\linewidth}\raggedright
B2 semantic
\end{minipage} \\
\midrule\noalign{}
\endhead
\bottomrule\noalign{}
\endlastfoot
Synthetic, non-adversarial (n=480) & φ −0.01 & φ 0.02 & --- & φ
\textbf{0.87} \\
Synthetic, adversarial (n=62) & admits 0.98 & --- & --- & catch
0.88→\textbf{0.48}; 37\% combined leak \\
Novel bespoke domain (n=124) & φ 0.00 & φ 0.09 & φ \textbf{0.41} (LOO) &
φ \textbf{0.02} \\
Real-world: retracted / predatory / farm (n=47) & φ 0.00 & φ −0.20 & ---
& φ 0.27; \textbf{0\%} of fraud flagged by panel \\
\end{longtable}
}

Three findings hold across the regimes. First, \textbf{structural
admission is statistically independent of truth} (φ = −0.01 on a
balanced 2×2, 95\% CI bracketing zero; odds ratio ≈ 1): the gate accepts
anchored content and rejects unanchored content, but whether an anchored
claim is \emph{true} is invisible to it, exactly the boundary §3 draws,
now measured rather than asserted, and a direct refutation of any
reading on which the three axioms make slop \emph{inadmissible}. Second,
\textbf{semantic judgment tracks truth only where the judge has domain
expertise}: a frontier judge that separated slop from legit at φ = 0.87
on generic content collapsed to φ = 0.02 on the novel CANONIC domain ---
it cannot tell real bespoke content from fabricated bespoke content ---
and an adversary optimizing against it halved its catch rate (0.88 →
0.48), with 37\% of adversarial slop passing structure and semantics
together; the most effective attacks were fluent-but-unsupported
mechanism, fabricated citations, and authoritative tone. Third,
\textbf{only grounding that verifies entailment against an evidence
corpus adds signal where expertise is absent}: retrieve-then-entail was
the single gate with positive truth-association in the novel-domain
regime and detected fabrication perfectly, but it is bound by corpus
coverage and cannot ground genuine novelty: under leave-one-out it
false-rejects 72\% of \emph{true} claims, because a corpus that excludes
a claim contains no independent support for it. (We flag an
evaluation-leakage artifact for replicators: scored naively, with the
held-out claims left in the index, the same gate reads φ = 1.0; the
honest figure is 0.41. Never evaluate a grounding gate on items drawn
from its own index.)

The real-world arm is the sharpest. On genuine
retracted-\emph{for-fraud} abstracts, the three-model panel labeled
\textbf{none} as slop and the semantic judge caught 17\%, because the
fraud is in the \emph{data}, not the \emph{prose}: the text reads as
legitimate science. The gates catch the slop that is easy to catch
(content-farm prose: 100\%) and miss the slop that is dangerous
(fabricated data, fluent fabrication). The conclusion the whole battery
supports is not that one gate beats the others but that \textbf{no
prose-reading gate reliably separates reliable from unreliable content},
because the highest-stakes unreliability is surface-legitimate and
genuine novelty cannot be grounded; both are verdicts only domain
expertise can render. What remains achievable --- and what CANONIC
delivers --- is \emph{accountability}: every admitted claim anchored to
a definition, a commit, and a window, so the slop verdict, which stays
with the expert, is rendered over a checkable ledger rather than over
unverifiable prose, and unreliable content cannot be unaccountable even
when it is wrong. The full harness, the pre-registration
(\texttt{PREREGISTRATION.md}, fixed before any run), and the per-regime
results are released at
\href{https://github.com/canonic-canonic/canonic-pub/tree/main/slop-benchmark}{\texttt{github.com/canonic-canonic/canonic-pub}}.

\subsubsection{Construction census}\label{construction-census}

The figures below count what the construction \emph{produced} and
confirm that it validates itself; they are a census, not an efficacy
measure. the benchmark above, not these counts, is the evidence about
what the gates filter. Table 6 reports the state at the window close,
the commit that declared the token economy on January 30, 2026, and each
figure is independently checkable against the public artifacts at that
ref. We report what existed when the construction window closed, not
what the corpus has grown into since; the difference between the two is
itself governed, and we return to it below.

\needspace{0.26\textheight}

\textbf{Table 6 \textbar{} Window-close metrics and how each is
independently verified at the closing ref.}

{\def\LTcaptype{none} 
\begin{longtable}[]{@{}
  >{\raggedright\arraybackslash}p{(\linewidth - 4\tabcolsep) * \real{0.2267}}
  >{\raggedright\arraybackslash}p{(\linewidth - 4\tabcolsep) * \real{0.1067}}
  >{\raggedright\arraybackslash}p{(\linewidth - 4\tabcolsep) * \real{0.6667}}@{}}
\toprule\noalign{}
\begin{minipage}[b]{\linewidth}\raggedright
Metric
\end{minipage} & \begin{minipage}[b]{\linewidth}\raggedright
Count
\end{minipage} & \begin{minipage}[b]{\linewidth}\raggedright
Verification
\end{minipage} \\
\midrule\noalign{}
\endhead
\bottomrule\noalign{}
\endlastfoot
Repositories & 10 & \texttt{ls\ -d\ */} across the orgs at the
window-close ref \\
Governed Scopes & 20 & \texttt{find\ .\ -name\ "CANON.md"} \\
Root Axioms & 3 & Triad, Inheritance, Introspection \\
Development Span & \textasciitilde1 month & December 29, 2025 ---
January 30, 2026 \\
Validation Status & PASS &
\texttt{python3\ validators/validator\_as\_a\_service.py} \\
\end{longtable}
}

Each metric has a reproducible verification method that any reader can
run after cloning the repositories and checking out the window-close
ref, since the framework resists claims that cannot be checked against
the artifact itself. A governed scope is simply a directory containing a
\texttt{CANON.md} file, so counting scopes reduces to a filesystem
operation rather than an interpretive judgement. The counts above are
the \emph{floors} of the construction: ten repositories and twenty
governed scopes had been declared by the time the token economy entered
governance. The evidence layer, session transcripts, episodes, and
invention disclosures, was declared in the window as governed scopes but
had barely begun to fill: at the closing ref one episode file and no
filed disclosures exist. That layer is the framework's mechanism for
accruing its own evidence, and it populated as the framework ran, as the
live-count discussion below records.

\subsubsection{Ecosystem composition}\label{ecosystem-composition}

The twenty window-close scopes are almost entirely \emph{core
governance}: the three axioms, the validators they require, the language
specification, and the induced analogs of an operating system, an
immutability layer, a distribution channel, and a token economy (Table
4). There are no archive scopes yet; archives are what a governed corpus
accumulates as surfaces retire, and at the window close nothing had
retired. This is the canonical point of the construction: a governance
framework is not a fixed inventory but a discipline for change. Run the
same \texttt{find\ .\ -name\ "CANON.md"} against the organizations at
\texttt{HEAD} today and the count is in the hundreds; every scope added
since resolves to its own governed commit, and the growth from twenty to
that figure is the framework governing its own expansion. We report
twenty as the window-close floor, and the live count as a separate,
later measurement, precisely so the two are never conflated.

The construction was documented contemporaneously in the CANONIC blog
series, itself a sequence of git-resolvable posts written as the work
happened: the December 29 genesis {[}17{]}, the nine-repository
milestone of mid-January {[}25{]}, and the velocity record we call
\emph{the exponential} {[}26{]}. We cite that series as the real-time
record, with one caveat we make explicit here because this paper is
fixed by a DOI while the posts are not. Written in the moment, several
posts report figures in rounded or narrative form, and a number of those
figures are the \emph{cumulative} totals of the still-growing corpus
reported as if they were the founding-window state; the same conflation
we correct above for the scope count, recurring across the
contemporaneous record for commit velocities and for organization and
repository tallies. Where any figure in that series diverges from the
git-resolved counts in this paper, evaluated at the window-close ref
(Appendix C), the counts here are authoritative. This paper is the
corrected record.

\subsubsection{Validation state}\label{validation-state}

Running the validator on the \texttt{canonic} repository at
\texttt{HEAD} returns the live count, grown past the twenty-scope
window-close floor of Table 6 as the corpus accreted:

\begin{verbatim}
=== VaaS - CANONIC Language Enforcement ===
Repository: canonic
Scopes found: 21
...
VALIDITY: PASS
\end{verbatim}

Every scope passes, meaning each one satisfies all three axioms: the
triad is present, inheritance is declared, and the vocabulary is closed.

Two further properties of the construction bear on its evaluation, and
we record them here rather than leave them implicit: the framework
validates itself, and it runs on commodity infrastructure.

\subsubsection{Self-application and the
Foundation}\label{self-application-and-the-foundation}

The \href{https://canonic.org/}{CANONIC Foundation}, a 501(c)(3)
nonprofit incorporated to host the governance framework documented in
this paper, operates under the same compiler-theoretic constraints it
enforces on downstream artifacts, a property we refer to as
self-application and which determines the institutional posture of the
organization at \textbf{canonic.org}. The artifacts produced under this
governance are themselves governed content subject to the same gates: a
clinical-evidence analysis of 77,388 trials posted as a preprint
{[}27{]}, and the long-form governed corpus that documents the framework
for its two principals, the developer and the governor {[}28{]}, are
both compiled through the validators described here, not exempt from
them. The framework is stewarded by the Foundation rather than by any
single operator or vendor, and the stewardship relation is itself a
governed scope whose CANON and validators are subject to the same
reproducibility, evidence-window, and lifecycle requirements imposed on
every other scope in the system.

Self-application has a precise technical reading. The manuscript that
describes CANONIC is itself a governed artifact: its CANON file declares
the constraints under which it is admissible, its validators are drawn
from the same registry used to admit any other manuscript, and its
distribution channels were discovered by the same channel-discovery
procedure the Foundation applies to unrelated downstream artifacts. The
system that the paper documents is therefore not merely described by the
paper but is, at the level of build and validation, the producer of the
paper, so that the closure between describing system and described
system is a structural property rather than a rhetorical one.

The economic loop that sustains the Foundation closes through the same
governed pipeline. Each session transcript enters the evidence layer as
a typed artifact; transcripts aggregate into episodes that satisfy the
evidence-window requirements of patent practice; episodes generate
invention disclosures that the Foundation files as provisional and
non-provisional applications; and the resulting intellectual property is
exchanged through the COIN ledger for the operating capital that funds
subsequent sessions. The cycle runs from transcript to episode to
disclosure to coin and back to transcript, and it is the mechanism by
which the framework underwrites its own continued development without
recourse to external custodianship.

\subsubsection{GitHub as substrate}\label{github-as-substrate}

The decision to host CANONIC governance artifacts on GitHub exposed a
latent property of the platform: a single substrate already provides
hosting, version-controlled immutability, continuous integration,
billing infrastructure via Sponsors, public verification surfaces, and
global distribution, six functions that together approximate the role of
an operating system for governed-content artifacts. The ten repositories
and twenty governed scopes produced during the construction window,
hosted under the canonic-machine and canonic-magic organizations, were
documented across the collaboration sessions against this substrate, and
an architectural review surfaced the observation as a structural rather
than incidental property of the deployment.

Figure 4 contrasts the components a comparable project would otherwise
have to construct against the components GitHub supplies out of the box.

\begin{figure}[!ht]
\centering

\pandocbounded{\includegraphics[keepaspectratio,alt={}]{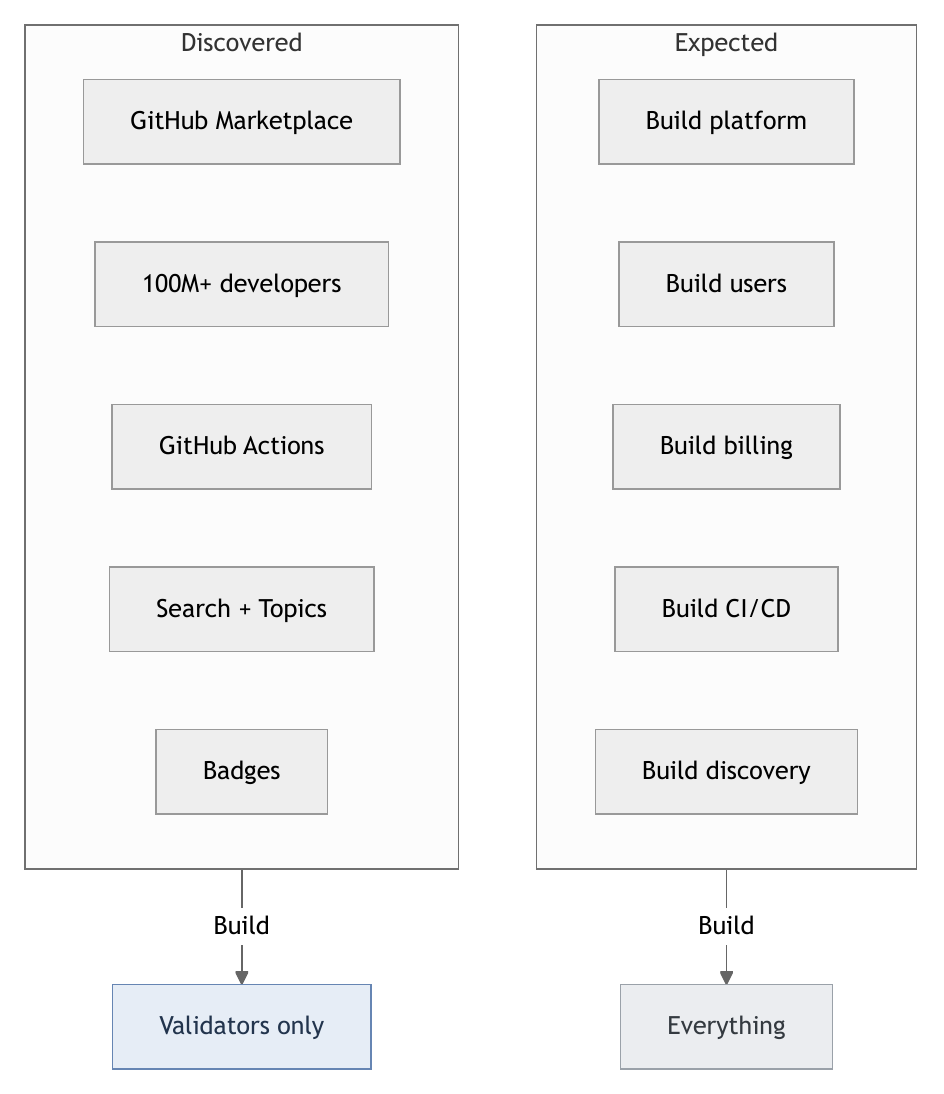}}

\textbf{Figure 4: Distribution layer realized as configuration over
existing infrastructure.} The \emph{Expected} column lists the
capabilities a typical project would need to build from scratch:
platform, users, billing, CI/CD, and discovery. The \emph{Discovered}
column lists the equivalent services GitHub already provides at no cost.
CANONIC builds only the remaining piece, \emph{Validation-as-a-Service}
validators, and treats the rest as configuration.

\end{figure}

The platform supplies hosting through the repository surface itself,
distribution through the Marketplace and topic-indexed search
infrastructure, continuous integration through GitHub Actions, billing
through GitHub Sponsors, an immutability guarantee through the
append-only commit graph that records every state transition, and a
public verification surface through the rendered repository view that
any reader can inspect without authentication. The residual component
required from CANONIC under this arrangement is therefore validation
alone, which the framework contributes as a reusable Action that
downstream repositories consume through a configuration file, as
illustrated by the following workflow snippet:

\begin{Shaded}
\begin{Highlighting}[]
\FunctionTok{name}\KeywordTok{:}\AttributeTok{ CANONIC Validation}
\FunctionTok{on}\KeywordTok{:}\AttributeTok{ }\KeywordTok{[}\AttributeTok{push}\KeywordTok{]}
\FunctionTok{jobs}\KeywordTok{:}
\AttributeTok{  }\FunctionTok{validate}\KeywordTok{:}
\AttributeTok{    }\FunctionTok{runs{-}on}\KeywordTok{:}\AttributeTok{ ubuntu{-}latest}
\AttributeTok{    }\FunctionTok{steps}\KeywordTok{:}
\AttributeTok{      }\KeywordTok{{-}}\AttributeTok{ }\FunctionTok{uses}\KeywordTok{:}\AttributeTok{ canonic{-}machine/vaas{-}action@v1}
\end{Highlighting}
\end{Shaded}

Repositories that pass validation render a status badge whose URL
resolves to the corresponding validator run, allowing third parties to
confirm compliance without local execution of the validator suite. The
arrangement collapses the boundary between supplemental material and
primary artifact: every repository cited in this manuscript is hosted
under the canonic-machine, canonic-magic, or canonic-canonic
organizations, per-repository visibility is governed by the LICENSE and
NOTICE files committed in each tree, every cited assertion is
reproducible by clone where licensing permits, and every validator
invocation is independently re-executable against the same commit graph
that the manuscript references.

The recursive property of the arrangement is that the manuscript
documenting the architecture was itself authored under CANONIC
governance, validated by the same validators it describes, and
distributed through the substrate it analyzes, a closure in which the
artifact, the evidence, and the verification surface coexist on a single
platform rather than across the conventional separation of paper,
supplement, and code repository. Treating GitHub as the substrate,
rather than as one of several deployment targets, removes the
engineering effort that would otherwise be required to assemble an
equivalent stack and reduces the trust surface for external verifiers to
a single well-understood platform with documented immutability and
access semantics.

\subsection{6 Related Work}\label{related-work}

Relocating verification to a gate at the point of admission has partial
precedents; CANONIC generalizes them by governing the content artifact
itself against a formal grammar at the corpus boundary. Table 7
positions it against the closest systems on five axes: what each gates,
when it acts, at what granularity, what audit trail it leaves, and where
control sits.

\needspace{0.26\textheight}

\textbf{Table 7 \textbar{} Where governance acts: CANONIC against
related gating and provenance systems.}

{\def\LTcaptype{none} 
\begin{longtable}[]{@{}
  >{\raggedright\arraybackslash}p{(\linewidth - 10\tabcolsep) * \real{0.1579}}
  >{\raggedright\arraybackslash}p{(\linewidth - 10\tabcolsep) * \real{0.1895}}
  >{\raggedright\arraybackslash}p{(\linewidth - 10\tabcolsep) * \real{0.1579}}
  >{\raggedright\arraybackslash}p{(\linewidth - 10\tabcolsep) * \real{0.1368}}
  >{\raggedright\arraybackslash}p{(\linewidth - 10\tabcolsep) * \real{0.2105}}
  >{\raggedright\arraybackslash}p{(\linewidth - 10\tabcolsep) * \real{0.1474}}@{}}
\toprule\noalign{}
\begin{minipage}[b]{\linewidth}\raggedright
System
\end{minipage} & \begin{minipage}[b]{\linewidth}\raggedright
What it gates
\end{minipage} & \begin{minipage}[b]{\linewidth}\raggedright
When
\end{minipage} & \begin{minipage}[b]{\linewidth}\raggedright
Granularity
\end{minipage} & \begin{minipage}[b]{\linewidth}\raggedright
Audit trail
\end{minipage} & \begin{minipage}[b]{\linewidth}\raggedright
Locus of control
\end{minipage} \\
\midrule\noalign{}
\endhead
\bottomrule\noalign{}
\endlastfoot
Open Policy Agent / HashiCorp Sentinel {[}8{]} & infrastructure and
config changes & pre-admission, at deploy time & per policy rule &
deployment-side decision logs & policy engine in the CI/CD path \\
C2PA {[}9{]} & media files & at capture or edit & per asset &
cryptographic manifest bound to the file & capture device or editor \\
Model and Data Cards {[}10{]} & disclosure about a model or dataset &
post-hoc, at release & per model or dataset & static document, no
enforced diff & author attestation \\
Study pre-registration {[}11{]} & an experimental protocol & pre-outcome
& per study & timestamped registry entry & an external registry \\
Anthropic Constitutional AI {[}12{]} & model outputs, as a training
signal & in training, pre-deployment & per response, absorbed into
weights & none; the constitution lives in the weights & the decoder \\
\textbf{CANONIC (this work)} & \textbf{the content artifact, against a
formal grammar} & \textbf{pre-admission, at the corpus boundary} &
\textbf{per claim: term, evidence, window} & \textbf{a git diff; every
constitution change is inspectable} & \textbf{the corpus boundary, not
the decoder} \\
\end{longtable}
}

Each prior system gates a single slice: a configuration change, a media
file, a disclosure, a protocol, a model's outputs. CANONIC differs in
kind on the last row. It governs the published artifact against a
versioned grammar, and control sits at the corpus boundary rather than
in the decoder. The system nearest in name, Constitutional AI {[}12{]},
is the farthest in mechanism: its constitution is a training signal
absorbed into weights that leaves no audit trail when it changes,
whereas CANONIC's constitution is a git diff anyone can inspect.

\subsection{7 Limitations}\label{limitations}

Several claims fall outside what this paper establishes, and we want to
mark them plainly rather than leave them to be inferred. The three
axioms we identify are sufficient for the domain we studied, and
Appendix B shows they are non-redundant, each is load-bearing (Theorem
5), but we do not show they are \emph{globally} minimal: we do not prove
that no smaller or differently shaped set would suffice. We chose them
because they emerged from operating practice across the evidence window,
not because we derived them from a proof of necessity. Other
configurations of axioms may turn out to govern other domains equally
well or better, and the work of finding them remains open.

CANONIC was built to govern governance specifications, and that is the
only domain we have actually run it on. Does the same
compile-and-enforce loop carry over to scientific datasets, legal
contracts, or software? We do not know. It is an empirical question and
we have not run the experiment. Scale is unproven too. Ten repositories
and twenty scopes at the window close is not thousands of repositories
and millions of scopes, and the engineering at that size will surface
problems we have not met. The specification is young: LANGUAGE.md sits
at version 0.1, it has known gaps, and we expect it to bend as new
domains push on its primitives.

The substrate is also single-node by construction. Everything reported
here ran on one developer's machine, with governance enforced at commit
time against local and GitHub-hosted repositories, the founding era,
which internally we call the \emph{canonic-machine} era after the
organization that held it. That era closed when its nine repositories
were sealed in late January, and its closing set the pattern for the
development model that followed: era-based, each subsequent era
extending the governed corpus rather than rewriting it, which is why the
live counts dwarf the window-close floor (§5). A distributed deployment
was planned from that point forward and is today realized as a
Docker-based distributed CANONIC, in which the same governance
primitives run as containerized services across nodes. That system is
deliberately out of scope here. This paper establishes the
compiler-theoretic primitives it is built on (the Triad, the inheritance
chain, the introspection contract, and the commit-time gate), and we
mark the boundary rather than cross it: nothing in the distributed
system is offered as evidence for the claims made here, and nothing here
depends on it.

Everything we report sits inside one window, December 29, 2025 to
January 30, 2026. Anything outside it is not evidence for these claims.

\subsection{8 Conclusion}\label{conclusion}

One move organizes the paper: treat admission as compilation --- a
mechanical, decidable decision about well-formedness made at the corpus
boundary, before publication, separate in kind from a judgment of truth
made after it. The three axioms (Triad, Inheritance, Introspection)
recover, through governance constraints rather than language design, the
structure of a compiler's syntax, scope-resolution, and type-system
layers, and they make admission a linear-time pass-or-fail rather than a
reviewer's verdict. A compiler does not prove a program correct. It
guarantees the program is well-formed enough to reason about. That,
precisely, is what CANONIC offers a corpus: governed intelligence that
compiles digital artifacts into an evidence ledger at scale. The
structural gate and the expert's verdict are not competing approaches.
They are complementary: the gate guarantees the record, and over that
record domain expertise renders the verdict.

We were careful to ask what that buys, and to answer it adversarially
rather than by assertion. A pre-registered, cross-provider benchmark
across non-adversarial synthetic, adversarial, novel-domain, and
real-world content returns a consistent and clarifying result:
structural admission is statistically independent of truth (φ ≈ 0),
retrieval-similarity grounding is likewise orthogonal to truth, semantic
judgment tracks truth only where the judge knows the domain and is
halved by fluent fabrication, and on real retracted-for-fraud abstracts
a frontier panel flags none. No prose-reading gate reliably keeps slop
out, because the dangerous failures (fabricated data, fluent
fabrication) are surface-legitimate, and genuine novelty cannot be
grounded against a corpus that does not yet contain it; both are
verdicts only domain expertise can render. The honest claim is therefore
not that CANONIC makes slop \emph{inadmissible} --- that verdict belongs
to the field, not to an algorithm; it is that CANONIC makes content
\emph{accountable}. Every admitted claim is anchored to a definition, a
commit, and an evidence window (entries in an append-only, git-anchored
ledger of work done that anyone can clone), so an unreliable artifact
cannot be \emph{unaccountable}: it can still be wrong, but it carries,
by construction, the provenance that lets domain expertise converge on
it, find it, contest it, and retract it.

Three limits bound what we can claim. First, the construction is n=1 and
self-authored, one coordinating author with the Claude Opus models (4.5,
then 4.7) over roughly a month; the benchmark restores independence to
the \emph{evaluation} (cross-provider judges, a non-author adversary,
validators rebuilt from the spec), but multi-author governance is still
untested. Second, the benchmark's adversaries are model families, not
human red-teams, its slop is largely synthetic, and grounding was
measured against a single corpus; the real-world arm is small. Third,
the three axioms are the set that fell out of this corpus, not a proven
floor, and a mechanized soundness proof for the validators remains
future work. The contribution we stand behind is the accountability
substrate and the measured map of where automated truth-filtering fails:
deciding what counts as slop is a verdict of domain expertise --- a
matter of taste, not a property an algorithm computes --- and CANONIC's
part is to maintain the source of truth that verdict is rendered over: a
distributed, git-anchored ledger of work done in which the evidence
behind every claim is reproducible, fully governed, and checkable end to
end.

\subsection{Author Context}\label{author-context}

This work emerges from two decades of research in precision medicine,
data annotation, and clinical AI {[}29--38{]}. The problem of AI
slop---content that sounds authoritative but lacks evidence---mirrors
the challenge of biomedical data annotation at scale {[}30, 32{]}.
Precision annotation of digital samples in NCBI's Gene Expression
Omnibus {[}30{]} required distinguishing validated labels from inferred
ones. Large-scale semi-automated labeling of clinical records {[}32{]}
required separating ground truth from approximation. The
governance/description separation in CANONIC applies the same principle:
what is LAW (CANON) versus what is DESCRIPTION (README).

Deep learning models for clinical diagnosis {[}34{]} surfaced a paradox
that has shadowed the field: the more powerful a model becomes the more
rigorous its validation must be if its outputs are to remain trustworthy
in clinical use. A model that predicts Alzheimer's disease from PET
scans {[}35{]} or estimates delirium risk in intensive care {[}34{]}
must trace every clinical claim back to the evidence that warrants it,
because diagnostic confidence in the absence of provenance is
indistinguishable from confabulation. CANONIC encodes that requirement
into the substrate itself, treating traceability not as a best practice
for careful teams but as a compilation requirement that every governed
artifact must satisfy before it can be admitted to the corpus.

The \href{https://github.com/canonic-machine/mammochat}{MammoChat
project} {[}31, 33, 39{]} is where this all started. It pioneered
patient-centered AI for breast cancer imaging during a period when most
clinical AI systems treated the patient as a data source rather than as
the principal whose interests the system was bound to serve. A \$2M
award from the Florida Department of Health's Casey DeSantis Cancer
Research Program funded it. Blockchain-based provenance ensured that
patients retained ownership of the data that the system used to reason
about them {[}31{]}, and that property became the seed for a broader
insight that CANONIC has since generalised. The framework extends the
same logic to all governed content by binding every claim to
cryptographic provenance through the underlying git substrate, with the
ledger serving as the immutable record of every governance transition.
CANONIC therefore emerged from practice rather than from theory, shaped
by the operational constraints of handling real patient data in a
setting where every claim carries clinical consequence.

\subsection{Acknowledgments}\label{acknowledgments}

The AI assistants that produced this work under governance are itemized
in the AI-assistance disclosure below, which records both the OpenAI
ChatGPT genesis conversation and the Anthropic Claude models of the
construction window. Their contributions are recorded in the ledger with
model-identity disclosure per CANONIC governance requirements.

The framework name honors the insight from \emph{Dividends \& Deaths}
where the governance/description separation first emerged---originally
as CANNON (with a typo), now as CANON.

\subsubsection{Funding}\label{funding}

This work builds on over \$5M in NIH-funded research spanning two
decades:

\textbf{\href{https://commonfund.nih.gov/bd2k}{BD2K (Big Data to
Knowledge)} Awards:} -
\href{https://reporter.nih.gov/search/?queryText=U01LM012675}{NIH
U01-LM012675}: ``Crowd-Assisted Deep Learning (CrADLe): Digital Curation
to Translate Big Data into Precision Medicine'' - NIH BD2K Crowdsourcing
Award, NCI (2016) -
\href{https://github.com/idrdex/stargeo}{STARGEO.org}: Search Tag
Analyze Resource for Gene Expression Omnibus

\textbf{Recognition:} - Inaugural Marcus Award for Precision Medicine
Innovation, UCSF (2016) - \textbf{\$2M} Casey DeSantis Cancer Research
Program award, Florida Department of Health, for MammoChat (2025) ---
the patient-centered breast-cancer navigation platform from which the
governance practice reported in this paper first emerged

\textbf{Training:} - MD/PhD in Genomics and Computational Biology,
University of Pennsylvania - Clinical Pathology Residency, Stanford
University - NIH-funded translational bioinformatics fellowship, UCSF
(Butte Lab)

\textbf{Institutional Support:} - CANONIC Foundation (Founder and Chair,
2026--present) - American Board of Precision Medicine (Director of AI,
2024--present)

\textbf{Former Institutional Positions} (where prior work cited here was
performed): - University of Central Florida, College of Medicine
(2019--2025) {[}MammoChat, clinical AI{]} - University of California,
San Francisco --- Assistant Professor (2015--2019)
{[}\href{https://github.com/hadleylab}{STARGEO}, CrADLe, BACPAC{]} -
Stanford University --- Engineering Research Associate, Butte Lab
(2013--2015) - Children's Hospital of Philadelphia --- Lead Clinical
Genomics Analyst (2010--2012) - University of Pennsylvania --- MD/PhD
(1999--2009)

The governance patterns crystallized here were forged in the constraints
of clinical AI where every claim must be defensible---from PennCNV
{[}36{]} to precision annotation {[}30{]} to breast cancer imaging
(Panahiazar {[}31, 33{]}). All funding is ledgered in NIH RePORTER. All
publications are ledgered in PubMed. CANONIC extends this ledger to
governance itself.

\subsection{Disclosures}\label{disclosures}

\textbf{Competing interests.}
\href{https://orcid.org/0000-0003-0990-4674}{Dexter Hadley} is Founder
and Chair of the non-profit CANONIC Foundation and Director of AI at the
American Board of Precision Medicine (ABOPM), a non-profit organization.
These are unpaid leadership roles --- the author receives no salary,
equity, or other compensation --- and are declared as non-financial
competing interests, since the work described in this manuscript is the
foundational thesis of the CANONIC Foundation, which the author leads.
The author declares no financial competing interest with the cited
references or with the AI vendors whose products this paper validates
against. CANONIC is offered as Apache License 2.0 open-source
governance; commercial follow-ons (VaaS --- Validators as a Service) are
tracked separately and are not cited in this paper as evidence.

\textbf{AI-assistance disclosure.} This manuscript was produced under
CANONIC governance, and the producing agent changed across the evidence
window in the way §4 describes. The December 29, 2025 genesis
conversation ran in OpenAI's ChatGPT (content-anchored in the Evidence
Ledger, § Context Anchors). The CANONIC framework itself was then
produced with the Anthropic Claude family from the January 5, 2026
kernel (\texttt{11affab}) forward: Claude Opus 4.5
(\texttt{claude-opus-4-5-20251101}) for v0.1 production through the
January 30, 2026 window close, and Claude Opus 4.7
(\texttt{claude-opus-4-7}) for v0.2 byline reconciliation against
governed VITAE (May 12, 2026). Model identity, session counts, and
turn-level transcripts are public in the supplemental LEDGER. v0.2
changed no claim or result; it reconciled affiliation and Institutional
Support against \texttt{hadleylab-canonic/USERS/DEXTER/VITAE.md} per
CANONIC's \texttt{byline\_forbidden\_phrases} constraint (UCF
Chief-of-AI title is durable historical, not a current institutional
affiliation). v0.3 (June 2026) extended the evidence window from the
21-day language-spec close (January 19) to the full construction month
ending January 30, 2026, so that the token-economy layer (COIN,
governance-declared January 30) sits inside the window rather than after
it; §5 counts are correspondingly reported as floors at the window
close, and no prior count or result changed. v0.4 (June 2026) adds the
§5 gate-efficacy benchmark --- a separate, pre-registered,
cross-provider evaluation declared as a \emph{later} measurement that
modifies no window-close count. Its judges and adversaries span three
model families (DeepSeek, OpenAI gpt-4o, Anthropic claude-sonnet) drawn
from a different family than the one that authored CANONIC, with a local
Apple-Silicon model for the perplexity baseline; the harness,
pre-registration, and per-regime results are released at
\href{https://github.com/canonic-canonic/canonic-pub/tree/main/slop-benchmark}{\texttt{github.com/canonic-canonic/canonic-pub}}.

\textbf{Validation status.} PASS --- every claim in this manuscript
resolves to a \texttt{LEDGER} row within the declared evidence window;
every governed scope cited passes the three CANONIC axioms (TRIAD,
INHERITANCE, INTROSPECTION); the manuscript itself was produced under
the same compilation gate it specifies. The one exception by
construction is the §5 benchmark (v0.4), whose claims resolve not to the
window \texttt{LEDGER} but to its separately-released harness and
per-regime results --- a later, independent evaluation rather than a
window-bounded construction claim.

\subsection{Code, Data, and Materials
Availability}\label{code-data-and-materials-availability}

\textbf{Code.} The CANONIC framework, validators, the LANGUAGE
specification, and the build toolchain are hosted at
\texttt{https://github.com/canonic-machine} and
\texttt{https://github.com/canonic-canonic}. The ten governed
repositories produced in the evidence window, the \texttt{bin/verify-*}
validator family, and the \texttt{LEDGER} schema are clone-and-run
reproducible where licensing permits; per-repository licensing is Apache
2.0 for the open core (e.g.~\texttt{canonic-machine/canonic},
\texttt{canonic-machine/PAPER}) and governed by the per-repo LICENSE and
NOTICE files for the remainder. Appendix C gives a step-by-step
reproducibility protocol; reviewers requiring access to
restricted-visibility repositories should contact the corresponding
author.

\textbf{Data.} This is a methodological / framework paper; there is no
separate dataset. The ``data'' is the LEDGER itself --- the session
transcripts, invention disclosures, and the full \texttt{git\ log} of
the window's ten repositories, all available at the same GitHub
footprint. The transcript and disclosure counts grow as the framework
runs; §5 reports the window-close structural totals at the closing ref,
and the live totals are recoverable by running the same commands at
\texttt{HEAD}. Every quantitative claim in §5 (Evidence Window) resolves
to a \texttt{find} / \texttt{git\ log} /
\texttt{python3\ validators/validator\_as\_a\_service.py} invocation
listed in Appendix D.

\textbf{Materials.} No physical or wet-lab materials. All materials are
digital governance artifacts (CANON.md, VOCAB.md, README.md, SPEC files,
LEDGER rows, validator binaries) hosted under the canonic-machine,
canonic-magic, and canonic-canonic organizations; per-repository
visibility is governed by the LICENSE and NOTICE files committed in each
tree.

\textbf{Models.} This work does not train or release a model. The AI
assistants used in production (Claude Opus 4.5, Claude Opus 4.7) are
commercial Anthropic models accessed via API; the LEDGER captures every
prompt, response, and tool-use turn for full reproducibility of the
human-AI collaboration trace.

\textbf{Pre-registration.} Not pre-registered --- this is a framework /
methodology paper, not a hypothesis-testing study. The roughly one-month
evidence window is the de facto declared scope.

\subsection{References}\label{references}

{[}1{]} Oxford Languages. (2025). Word of the Year 2025: ``AI Slop.''
Oxford University Press. URL:
https://languages.oup.com/word-of-the-year/2025/

{[}2{]} Vaswani, A., et al.~(2017).
\href{https://arxiv.org/abs/1706.03762}{Attention Is All You Need}.
\emph{NeurIPS 2017}, 5998--6008. arXiv:1706.03762. URL:
https://arxiv.org/abs/1706.03762

{[}3{]} Brown, T., et al.~(2020).
\href{https://arxiv.org/abs/2005.14165}{Language Models are Few-Shot
Learners}. \emph{NeurIPS 2020}, 1877--1901. arXiv:2005.14165. URL:
https://arxiv.org/abs/2005.14165

{[}4{]} Weidinger, L., et al.~(2021). Ethical and social risks of harm
from Language Models. \emph{arXiv preprint}. arXiv:2112.04359. URL:
https://arxiv.org/abs/2112.04359

{[}5{]} Ji, Z., et al.~(2023). Survey of Hallucination in Natural
Language Generation. \emph{ACM Computing Surveys}, 55(12), 1--38. DOI:
10.1145/3571730

{[}6{]} Bender, E.M., Gebru, T., McMillan-Major, A., \& Shmitchell, S.
(2021). On the Dangers of Stochastic Parrots: Can Language Models Be Too
Big? \emph{FAccT '21}, 610--623. DOI: 10.1145/3442188.3445922

{[}7{]} Backus, J.W., et al.~(1960).
\href{https://doi.org/10.1145/367236.367262}{Report on the Algorithmic
Language ALGOL 60}. \emph{Communications of the ACM}, 3(5), 299--314.
DOI: 10.1145/367236.367262

{[}8{]} Open Policy Agent. (2021).
\href{https://www.openpolicyagent.org/}{Policy-based control for
cloud-native environments}. Cloud Native Computing Foundation (graduated
project). URL: https://www.openpolicyagent.org/ (cf.~HashiCorp Sentinel,
https://docs.hashicorp.com/sentinel)

{[}9{]} Coalition for Content Provenance and Authenticity (C2PA).
(2024). \href{https://c2pa.org/specifications/}{Technical Specification
v2.1}. URL: https://c2pa.org/specifications/

{[}10{]} Mitchell, M., Wu, S., Zaldivar, A., et al.~(2019).
\href{https://doi.org/10.1145/3287560.3287596}{Model Cards for Model
Reporting}. \emph{FAT} '19\emph{, 220--229. DOI: 10.1145/3287560.3287596
(cf.~Gebru, T., et al.~(2021). Datasheets for Datasets. }CACM* 64(12),
86--92. DOI: 10.1145/3458723)

{[}11{]} Nosek, B.A., Ebersole, C.R., DeHaven, A.C., \& Mellor, D.T.
(2018). \href{https://doi.org/10.1073/pnas.1708274114}{The
preregistration revolution}. \emph{PNAS}, 115(11), 2600--2606. DOI:
10.1073/pnas.1708274114

{[}12{]} Bai, Y., Kadavath, S., Kundu, S., et al.~(2022).
\href{https://arxiv.org/abs/2212.08073}{Constitutional AI: Harmlessness
from AI Feedback}. \emph{arXiv preprint}. arXiv:2212.08073. URL:
https://arxiv.org/abs/2212.08073

{[}13{]} Hadley, D. (2026).
\href{https://hadleylab.org/blogs/2026-04-03-constitutional-ai-vs-canonic}{Constitutional
AI Is Not a Constitution}. CANONIC governance blog. URL:
https://hadleylab.org/blogs/2026-04-03-constitutional-ai-vs-canonic
(delineates Anthropic's model-scoped Constitutional AI from CANONIC's
institution-scoped governance).

{[}14{]} Chomsky, N. (1959). On Certain Formal Properties of Grammars.
\emph{Information and Control}, 2(2), 137--167. DOI:
10.1016/S0019-9958(59)90362-6

{[}15{]} Knuth, D.E. (1965). On the Translation of Languages from Left
to Right. \emph{Information and Control}, 8(6), 607--639. DOI:
10.1016/S0019-9958(65)90426-2

{[}16{]} Aho, A.V., Lam, M.S., Sethi, R., \& Ullman, J.D. (2006).
\emph{Compilers: Principles, Techniques, and Tools} (2nd ed.).
Addison-Wesley. ISBN: 0-321-48681-1. URL:
https://suif.stanford.edu/dragonbook/

{[}17{]} Hadley, D. (2025).
\href{https://hadleylab.org/blogs/2025-12-29-the-compiler-insight}{The
Compiler Insight}. CANONIC governance blog, December 29, 2025. URL:
https://hadleylab.org/blogs/2025-12-29-the-compiler-insight (origin of
the governance-is-compilation reframing).

{[}18{]} Chomsky, N. (1956). Three Models for the Description of
Language. \emph{IRE Transactions on Information Theory}, 2(3), 113--124.
DOI: 10.1109/TIT.1956.1056813

{[}19{]} Johnston, J.B. (1971). The Contour Model of Block Structured
Processes. \emph{ACM SIGPLAN Notices}, 6(2), 55--82. DOI:
10.1145/942582.807990

{[}20{]} Milner, R. (1978). A Theory of Type Polymorphism in
Programming. \emph{Journal of Computer and System Sciences}, 17(3),
348--375. DOI: 10.1016/0022-0000(78)90014-4

{[}21{]} Ritchie, D.M. (1993).
\href{https://doi.org/10.1145/154766.155580}{The Development of the C
Language}. \emph{History of Programming Languages II}, 671--698. DOI:
10.1145/154766.155580

{[}22{]} Gosling, J., et al.~(2021).
\href{https://docs.oracle.com/javase/specs/jls/se17/html/index.html}{\emph{The
Java Language Specification} (Java SE 17 ed.)}. Oracle. URL:
https://docs.oracle.com/javase/specs/jls/se17/html/index.html

{[}23{]} Bradbury, S., et al.~(2025). \href{https://go.dev/ref/spec}{The
Go Programming Language Specification}. URL: https://go.dev/ref/spec

{[}24{]} Klabnik, S., \& Nichols, C. (2023).
\href{https://doc.rust-lang.org/book/}{\emph{The Rust Programming
Language}}. No Starch Press. URL: https://doc.rust-lang.org/book/

{[}25{]} Hadley, D. (2026).
\href{https://hadleylab.org/blogs/2026-01-15-nine-repos}{Nine Repos}.
CANONIC governance blog, January 15, 2026. URL:
https://hadleylab.org/blogs/2026-01-15-nine-repos (the canonic-machine
founding era at its nine-repository milestone; a git-resolvable
contemporaneous artifact).

{[}26{]} Hadley, D. (2026).
\href{https://hadleylab.org/blogs/2026-02-11-the-exponential}{The
Exponential}. CANONIC governance blog, February 11, 2026. URL:
https://hadleylab.org/blogs/2026-02-11-the-exponential (the
governance-compounding velocity record of the founding era; cumulative
figures therein are superseded by the git-resolved window-close counts
in §5).

{[}27{]} Bajnath, A., Haraksingh, R., Forghani, I., Evans, A., Cyrus,
E., Nimrod, M., \& Hadley, D. (2026).
\href{https://doi.org/10.64898/2026.05.14.26353197}{Enter the Matrix of
Drug Discovery: Race, Ethnicity, and the Evidence Gap Across 77,388
Clinical Trials}. \emph{medRxiv}. DOI: 10.64898/2026.05.14.26353197 (a
governed manuscript produced under the framework described here).

{[}28{]} Hadley, D. (2026). The CANONIC Doctrine (developer's
perspective) and The CANONIC Canon (governor's perspective). CANONIC
Foundation governed long-form. URL:
https://hadleylab.org/books/CANONIC-DOCTRINE/ ·
https://hadleylab.org/books/CANONIC-CANON/

{[}29{]} Hadley, D., et al.~(2017).
\href{https://pubmed.ncbi.nlm.nih.gov/28936969/}{Systematic integration
of biomedical knowledge prioritizes drugs for repurposing}.
\emph{eLife}, 6, e26726. PMID: 28936969.

{[}30{]} Hadley, D., et al.~(2017).
\href{https://pubmed.ncbi.nlm.nih.gov/28925997/}{Precision annotation of
digital samples in NCBI's gene expression omnibus}. \emph{Scientific
Data}, 4, 170125. PMID: 28925997.

{[}31{]} Panahiazar, M., Chen, N., Lituiev, D., \& Hadley, D. (2022).
\href{https://pubmed.ncbi.nlm.nih.gov/34697751/}{Empowering study of
breast cancer data with application of artificial intelligence
technology}. \emph{Clinical \& Experimental Metastasis}, 39(1),
117--127. PMID: 34697751.

{[}32{]} Ding, S., et al.~(2019).
\href{https://pubmed.ncbi.nlm.nih.gov/30128778/}{Large Scale
Semi-Automated Labeling of Routine Free-Text Clinical Records for Deep
Learning}. \emph{Journal of Digital Imaging}, 32(1), 30--37. PMID:
30128778.

{[}33{]} Ding, S., et al.~(2019). Automatic Labeling of Special
Diagnostic Mammography Views from Images and DICOM Headers.
\emph{Journal of Digital Imaging}, 32(2), 228--233. PMID: 30465142.

{[}34{]} Wong, W., et al.~(2018).
\href{https://pubmed.ncbi.nlm.nih.gov/30646095/}{Development and
Validation of an Electronic Health Record-Based Machine Learning Model
to Estimate Delirium Risk}. \emph{JAMA Network Open}, 1(4), e181018.
PMID: 30646095.

{[}35{]} Ding, S., et al.~(2019).
\href{https://pubmed.ncbi.nlm.nih.gov/30398430/}{A Deep Learning Model
to Predict a Diagnosis of Alzheimer Disease by Using 18F-FDG PET of the
Brain}. \emph{Radiology}, 290(2), 456--464. PMID: 30398430.

{[}36{]} Wang, K., Li, M., Hadley, D., et al.~(2007).
\href{https://pubmed.ncbi.nlm.nih.gov/17921354/}{PennCNV: An integrated
hidden Markov model designed for high-resolution copy number variation
detection}. \emph{Genome Research}, 17(11), 1665--1674. PMID: 17921354.

{[}37{]} Hadley, D., et al.~(2020). The Impact of COVID-19 on African
American Communities in the United States. \emph{Health Equity}, 4(1),
476--483. PMID: 33269331.

{[}38{]} Gianfrancesco, M., et al.~(2019). Tracing diagnosis
trajectories over millions of patients reveal an unexpected risk in
schizophrenia. \emph{Scientific Data}, 6, 210. PMID: 31615985.

{[}39{]} MammoChat. (2025). Empowering Women with Empathic AI. URL:
https://github.com/canonic-machine/mammochat

{[}40{]} Hindley, R. (1969).
\href{https://doi.org/10.1090/S0002-9947-1969-0253905-6}{The Principal
Type-Scheme of an Object in Combinatory Logic}. \emph{Transactions of
the American Mathematical Society}, 146, 29--60. DOI:
10.1090/S0002-9947-1969-0253905-6

{[}41{]} Dijkstra, E.W. (1960).
\href{https://doi.org/10.1007/BF01386232}{Recursive Programming}.
\emph{Numerische Mathematik}, 2(1), 312--318. DOI: 10.1007/BF01386232

{[}42{]} Landin, P.J. (1964).
\href{https://doi.org/10.1093/comjnl/6.4.308}{The Mechanical Evaluation
of Expressions}. \emph{The Computer Journal}, 6(4), 308--320. DOI:
10.1093/comjnl/6.4.308

{[}43{]} Cardelli, L., \& Wegner, P. (1985).
\href{https://doi.org/10.1145/6041.6042}{On Understanding Types, Data
Abstraction, and Polymorphism}. \emph{ACM Computing Surveys}, 17(4),
471--523. DOI: 10.1145/6041.6042

{[}44{]} Pierce, B.C. (2002).
\href{https://mitpress.mit.edu/9780262162098/}{\emph{Types and
Programming Languages}}. MIT Press. ISBN: 0-262-16209-1. URL:
https://mitpress.mit.edu/9780262162098/

{[}45{]} Curry, H.B., \& Feys, R. (1958).
\href{https://archive.org/details/combinatorylogic0001curr}{\emph{Combinatory
Logic, Volume I}}. North-Holland. URL:
https://archive.org/details/combinatorylogic0001curr

{[}46{]} Howard, W.A. (1980). The Formulae-as-Types Notion of
Construction. In \emph{To H.B. Curry: Essays on Combinatory Logic},
479--490. Academic Press. ISBN: 0-12-490200-3

{[}47{]} Martin-Löf, P. (1984).
\href{https://archive-pml.github.io/martin-lof/pdfs/Bibliopolis-Book-1984.pdf}{\emph{Intuitionistic
Type Theory}}. Bibliopolis. URL:
https://archive-pml.github.io/martin-lof/pdfs/Bibliopolis-Book-1984.pdf

{[}48{]} Coquand, T., \& Huet, G. (1988).
\href{https://doi.org/10.1016/0890-5401(88)90005-3}{The Calculus of
Constructions}. \emph{Information and Computation}, 76(2--3), 95--120.
DOI: 10.1016/0890-5401(88)90005-3

{[}49{]} Lamport, L. (1978).
\href{https://doi.org/10.1145/359545.359563}{Time, Clocks, and the
Ordering of Events in a Distributed System}. \emph{Communications of the
ACM}, 21(7), 558--565. DOI: 10.1145/359545.359563

{[}50{]} Merkle, R.C. (1988).
\href{https://doi.org/10.1007/3-540-48184-2_32}{A Digital Signature
Based on a Conventional Encryption Function}. \emph{CRYPTO '87
Proceedings}, LNCS 293, 369--378. DOI: 10.1007/3-540-48184-2\_32

{[}51{]} Nakamoto, S. (2008). Bitcoin: A Peer-to-Peer Electronic Cash
System. URL: https://bitcoin.org/bitcoin.pdf

{[}52{]} Torvalds, L., \& Hamano, J. (2005). Git: Fast Version Control
System. URL: https://git-scm.com/

{[}53{]} Anthropic. (2025--2026). Claude Opus model family
(\texttt{claude-opus-4-5}, \texttt{claude-opus-4-7}) --- model
documentation. URL: https://www.anthropic.com/claude

{[}54{]} OpenAI. (2023). \href{https://arxiv.org/abs/2303.08774}{GPT-4
Technical Report}. \emph{arXiv preprint}. arXiv:2303.08774. URL:
https://arxiv.org/abs/2303.08774

{[}55{]} Ostrom, E. (1990).
\href{https://doi.org/10.1017/CBO9780511807763}{\emph{Governing the
Commons: The Evolution of Institutions for Collective Action}}.
Cambridge University Press. ISBN: 0-521-40599-8. DOI:
10.1017/CBO9780511807763

{[}56{]} Lessig, L. (1999).
\href{https://archive.org/details/codeotherlawsofc00less}{\emph{Code and
Other Laws of Cyberspace}}. Basic Books. ISBN: 0-465-03913-8. URL:
https://archive.org/details/codeotherlawsofc00less

{[}57{]} Wright, A., \& De Filippi, P. (2015).
\href{https://papers.ssrn.com/sol3/papers.cfm?abstract_id=2580664}{Decentralized
Blockchain Technology and the Rise of Lex Cryptographia}. \emph{SSRN
Electronic Journal}. SSRN: 2580664. URL:
https://papers.ssrn.com/sol3/papers.cfm?abstract\_id=2580664

{[}58{]} Hadley, D. (2026). CANONIC LANGUAGE Specification v0.1. URL:
https://github.com/canonic-machine/canonic

{[}59{]} Hadley, D. (2026). CANONIC COVERAGE Specification. URL:
https://github.com/canonic-machine/canonic/blob/main/COVERAGE.md

{[}60{]} Hadley, D. (2026). CANONIC VOCAB Definitions. URL:
https://github.com/canonic-machine/canonic/blob/main/VOCAB.md

{[}61{]} Hadley, D. (2026). VaaS: Validators as a Service. URL:
https://github.com/canonic-machine/VALIDATORS

{[}62{]} Hadley, D. (2026). CANONIC CANON (Root Axioms). URL:
https://github.com/canonic-machine/canonic/blob/main/CANON.md

{[}63{]} STARGEO. (2016). Search Tag Analyze Resource for Gene
Expression Omnibus. URL: https://github.com/idrdex/stargeo

{[}64{]} Hadley Lab. (2025). Translating Big Data into Precision
Medicine. URL: https://github.com/hadleylab

{[}65{]} CANONIC Foundation (operates VaaS validators). (2026).
Constitutional AI Governance Framework. URL: https://canonic.org/ \#
Appendix A: Root Axioms

Verbatim from \texttt{canonic/CANON.md}:

\textbf{Axiom 0 --- Triad}

\begin{quote}
A scope \textbf{MUST} contain: \texttt{CANON.md}, \texttt{VOCAB.md},
\texttt{README.md}.
\end{quote}

\textbf{Axiom 1 --- Inheritance}

\begin{quote}
Every \texttt{CANON.md} \textbf{MUST} declare \texttt{inherits:}.
Inheritance chains \textbf{MUST} terminate at \texttt{/}. Inherited
axioms are final.
\end{quote}

\textbf{Axiom 2 --- Introspection}

\begin{quote}
\texttt{VOCAB.md} \textbf{MUST} define every concept used in
\texttt{CANON.md} and \texttt{VOCAB.md}.
\end{quote}

\section{Appendix B: Formal
Foundations}\label{appendix-b-formal-foundations}

\subsection{B.1 Definitions}\label{b.1-definitions}

The formal definitions below equip CANONIC's three axioms, namely
\emph{Triad}, \emph{Inheritance}, and \emph{Introspection}, with the
precise vocabulary required for the independence and minimality results
in §B.3 and the decidability proof in §B.4.

\textbf{Definition 1 (Scope).} A scope \(S\) is a tuple \((P, T, A, V)\)
where: - \(P\) is a filesystem path, - \(T = \{\)\texttt{CANON.md},
\texttt{VOCAB.md}, \texttt{README.md}\(\}\) is the triad, - \(A\) is a
set of axioms declared in \texttt{CANON.md}, - \(V\) is a set of
definitions declared in \texttt{VOCAB.md}.

\textbf{Definition 2 (Inheritance Chain).} For a scope \(S\) with path
\(P\), the inheritance chain \(I(S)\) is the sequence

\[I(S) = [S, \mathrm{parent}(S), \mathrm{parent}(\mathrm{parent}(S)), \ldots, \mathrm{root}],\]

where \(\mathrm{parent}(S)\) denotes the scope rooted at the parent
directory of \(P\), and \(\mathrm{root}\) denotes the scope at path
\texttt{/}.

\textbf{Definition 3 (Vocabulary Closure).} A scope \(S\) is
\emph{vocabulary-closed} if and only if

\[\forall\, t \in \mathrm{terms}(\mathrm{CANON}_S) \cup \mathrm{terms}(\mathrm{VOCAB}_S) : t \in \mathrm{definitions}(\mathrm{VOCAB}_S) \cup \mathrm{definitions}(I(S)),\]

so that every term appearing in either \texttt{CANON.md} or
\texttt{VOCAB.md} is defined either locally or within some ancestor
scope along the inheritance chain.

\textbf{Definition 4 (Validity).} A scope \(S\) is \emph{valid} if and
only if the triad \(T\) is contained in \(\mathrm{files}(P)\), the
inheritance chain \(I(S)\) terminates at \(\mathrm{root}\), and \(S\)
satisfies the vocabulary-closure condition of Definition 3.

\subsection{B.2 Compiler
Correspondence}\label{b.2-compiler-correspondence}

\textbf{Theorem 1 (Syntactic Correspondence).} The \emph{triad} axiom
corresponds to the syntactic well-formedness requirements of a
context-free grammar in the sense of Chomsky {[}14{]} and Backus
{[}7{]}.

\emph{Proof.} A context-free grammar G = (V, Σ, R, S) requires: - V: set
of non-terminal symbols - Σ: set of terminal symbols - R: set of
production rules - S: start symbol

We construct a mapping φ: CANONIC → CFG: - φ(scope identifiers) = V -
φ(\{\texttt{CANON.md}, \texttt{VOCAB.md}, \texttt{README.md}\}) = Σ -
φ(triad requirement) = R, specifically: Scope → CANON VOCAB README -
φ(root) = S

The triad axiom mandates that for every valid scope:

\begin{verbatim}
files(S) ⊇ {CANON.md, VOCAB.md, README.md}
\end{verbatim}

This constraint is a syntactic well-formedness condition in the
grammatical sense, since a scope is syntactically valid if and only if
it contains the required terminal symbols, and the mapping φ preserves
the condition by construction. The correspondence is narrow, and we
state its limit plainly: the production R = (Scope → CANON VOCAB README)
is a single non-recursive rule, so φ exhibits a finite well-formedness
checklist with the \emph{form} of a context-free production, not a
context-free language. The triad is the degenerate case of a grammar,
which is exactly why its admission check is a set-containment test.

Therefore, the \emph{triad} axiom corresponds to syntactic
well-formedness in this restricted sense. □

\textbf{Theorem 2 (Scope Resolution Correspondence).} The
\emph{inheritance} axiom implements static scope resolution equivalent
to the contour model of Johnston {[}19{]}.

\emph{Proof.} The contour model {[}19{]} defines name resolution in
block-structured languages by the recurrence:

\begin{verbatim}
resolve(name, block) =
    if name ∈ declarations(block) then lookup(name, block)
    else resolve(name, enclosing(block))
\end{verbatim}

The inheritance axiom defines concept resolution by the parallel
recurrence:

\begin{verbatim}
resolve(term, scope) =
    if term ∈ definitions(VOCAB_scope) then definition(term, scope)
    else resolve(term, parent(scope))
\end{verbatim}

The structural correspondence between the two procedures is evident
under the substitutions block ↔ scope, declarations ↔
definitions(VOCAB), enclosing ↔ parent, and name ↔ term. Both procedures
terminate; the contour model halts at the outermost enclosing block, and
CANONIC halts at the root scope, with root termination guaranteed by
Axiom 1, which requires that every inheritance chain terminate at ``/''.

Therefore, the \emph{inheritance} axiom corresponds to lexical scope
resolution as formalized by the contour model. □

\textbf{Theorem 3 (Type System Correspondence).} The
\emph{introspection} axiom implements a closed system of vocabulary that
is equivalent, in its closure property, to the completeness condition of
the Hindley-Milner type system due to Milner {[}20{]} (see also Hindley
1969 {[}40{]} and Aho et al.~{[}16{]} for the compiler-construction
perspective).

\emph{Proof.} The Hindley-Milner type system possesses the principal
type property, under which every well-typed expression admits a most
general type inferable without annotation, and the load-bearing
structural property of the system is \emph{closure}: every type
reference within a well-typed program resolves to a type definition
within the same program.

Introspection requires the analogous closure condition:

\begin{verbatim}
∀ t ∈ terms(CANON) ∪ terms(VOCAB) : t ∈ definitions(VOCAB*)
\end{verbatim}

where VOCAB* denotes the transitive closure of VOCAB through the
inheritance chain.

This condition expresses vocabulary closure: no undefined term may
appear within a valid scope. The correspondence between the two systems
is given by the substitutions type reference ↔ term usage, type
definition ↔ vocabulary definition, type inference ↔ definition lookup
through the inheritance chain, and principal types ↔ canonical
definitions, since the closest definition along the inheritance chain is
final and unique.

A CANONIC scope is therefore well-typed if and only if every term
resolves to exactly one definition through inheritance, which parallels
the manner in which Hindley-Milner infers the principal type of every
well-typed expression.

Therefore, the \emph{introspection} axiom is isomorphic to a closed
\emph{type system} in the sense of Hindley {[}40{]} and Milner {[}20{]}.
□

\subsection{B.3 Independence and
Minimality}\label{b.3-independence-and-minimality}

We establish two structural properties of the three axioms (that they
are mutually \emph{independent} and individually \emph{minimal}), and
are explicit about a third property we do \textbf{not} claim. An earlier
formulation of this work asserted a \emph{completeness} result: that AI
slop is structurally inadmissible under the three axioms. The §5
benchmark refutes that assertion directly. Structural admission is
statistically independent of content-truth (φ ≈ 0), so the axioms bound
\emph{accountability}, not truth, and no closure property of the triad
makes slop inadmissible. We therefore retain the independence and
minimality theorems below, which characterize the axiom set, and
withdraw the completeness claim.

\textbf{Theorem 4 (Axiom Independence).} The three axioms are mutually
independent.

\emph{Proof.} We construct counterexamples showing that each axiom can
fail while the other two continue to hold:

\begin{enumerate}
\def\labelenumi{\arabic{enumi}.}
\item
  \emph{Triad fails, others hold:} a directory containing only
  \texttt{CANON.md} and \texttt{VOCAB.md}, where CANON declares
  inheritance to root and VOCAB defines every term in use. The triad is
  incomplete, yet inheritance and introspection are satisfied.
\item
  \emph{Inheritance fails, others hold:} a scope with a complete triad
  and closed vocabulary in which CANON declares
  \texttt{inherits:\ /nonexistent/path/}, so the chain fails to
  terminate at root despite the other two properties holding.
\item
  \emph{Introspection fails, others hold:} a scope with a complete triad
  and a valid inheritance chain in which CANON uses the term ``foobar''
  that appears in no VOCAB along the chain.
\end{enumerate}

Because each axiom can fail in isolation while the remaining two are
satisfied, no axiom is derivable from the others. □

\textbf{Theorem 5 (Axiom Minimality).} Removing any single axiom breaks
the governance system.

\emph{Proof.} We consider each removal in turn:

\begin{enumerate}
\def\labelenumi{\arabic{enumi}.}
\item
  \emph{Remove Triad:} without mandatory files there is no structure
  left to validate, so a directory containing arbitrary files could
  claim validity and governance becomes undefined.
\item
  \emph{Remove Inheritance:} without grounded authority chains, scopes
  are free to claim arbitrary parents or to form cycles, and authority
  becomes unverifiable in the general case.
\item
  \emph{Remove Introspection:} without vocabulary closure, terms can
  mean anything the author wishes, so slop becomes admissible whenever
  undefined jargon passes validation.
\end{enumerate}

Each removal admits a class of invalid content that the full three-axiom
system rejects. □

\textbf{Corollary (Compiler Correspondence Coverage).} The three axioms
together cover the three fundamental compiler concepts, relative to the
standard pedagogical decomposition.

\emph{Proof.} By Theorems 1 through 3 we have the correspondences: -
\emph{triad} ≅ \emph{syntax} (structure) - \emph{inheritance} ≅
\emph{scope resolution} (binding) - \emph{introspection} ≅ \emph{type
system} (semantics)

By Theorems 4 and 5 the axioms are independent and minimal. Syntax,
scope resolution, and type systems are, in the standard pedagogical
decomposition, the three pillars of classical compiler theory {[}16{]}.
CANONIC supplies governance analogs for each, and only for those three,
so the correspondence is complete \emph{relative to that decomposition}
--- a qualification, not a claim that no other partition of compiler
theory exists. □

\subsection{B.4 Decidability}\label{b.4-decidability}

\textbf{Theorem 6 (Validation Decidability).} Scope validity is
decidable in O(n) where n is the total size of all files in the
inheritance chain.

\emph{Proof.} The validation algorithm decomposes into three phases
whose decidability follows in turn from filesystem existence,
bounded-depth chain traversal with cycle detection, and finite set-union
with subset testing:

\begin{verbatim}
validate(scope):
    // Triad check: O(1)
    if not exists(CANON.md) or not exists(VOCAB.md) or not exists(README.md):
        return INVALID

    // Inheritance check: O(d) where d = chain depth
    chain = []
    current = scope
    while current != root:
        if current in chain:  // cycle detection
            return INVALID
        chain.append(current)
        current = parent(current)

    // Introspection check: O(n)
    all_definitions = union(definitions(v) for v in chain)
    all_terms = union(terms(c) for c in chain) ∪ union(terms(v) for v in chain)
    if not all_terms ⊆ all_definitions:
        return INVALID

    return VALID
\end{verbatim}

The cost of the three phases sums to O(1) + O(d) + O(n) = O(n) under d ≤
n, so the procedure halts on every input. □

\section{Appendix C: Reproducibility
Protocol}\label{appendix-c-reproducibility-protocol}

\begin{Shaded}
\begin{Highlighting}[]
\CommentTok{\# Clone the evidence and PIN to the evidence{-}window close.}
\CommentTok{\# The window (Dec 29 2025 – Jan 30 2026) is fixed. Every count below is}
\CommentTok{\# evaluated at the last commit dated on or before the closing day, not at}
\CommentTok{\# mutable HEAD, so the figures are reproducible and the declared window cannot}
\CommentTok{\# be silently re{-}chosen after the fact (cf. the window{-}shopping attack, §3).}
\FunctionTok{git}\NormalTok{ clone https://github.com/canonic{-}machine/canonic.git}
\BuiltInTok{cd}\NormalTok{ canonic}
\VariableTok{WINDOW\_REF}\OperatorTok{=}\StringTok{"}\VariableTok{$(}\FunctionTok{git}\NormalTok{ rev{-}list }\AttributeTok{{-}1} \AttributeTok{{-}{-}until}\OperatorTok{=}\NormalTok{2026{-}01{-}30 HEAD}\VariableTok{)}\StringTok{"}   \CommentTok{\# the SHA the claims resolve to}
\FunctionTok{git}\NormalTok{ checkout }\StringTok{"}\VariableTok{$WINDOW\_REF}\StringTok{"}

\CommentTok{\# Run validators}
\ExtensionTok{python3}\NormalTok{ validators/validator\_as\_a\_service.py}
\CommentTok{\# Expected: VALIDITY: PASS}

\CommentTok{\# Count governed scopes (CANON.md files) at the window{-}close ref}
\FunctionTok{find}\NormalTok{ . }\AttributeTok{{-}name} \StringTok{"CANON.md"} \KeywordTok{|} \FunctionTok{wc} \AttributeTok{{-}l}

\CommentTok{\# Verify a specific pinned artifact}
\FunctionTok{git}\NormalTok{ show 11affab}
\CommentTok{\# Expected: first CANON.md content}
\end{Highlighting}
\end{Shaded}

Each repository's window-close \texttt{WINDOW\_REF} is the SHA against
which its claims in §5 are stated, and the command above recomputes it
deterministically from the public history, so no separate manifest is
trusted. The counts in §5 are reported as floors at the window close:
the construction was still active through January 30, so a later ref can
only equal or exceed them. \textbf{If a count evaluated at
\texttt{WINDOW\_REF} is fewer than the value reported in §5, the
corresponding claim is falsified.}

\section{Appendix D: Evidence Index}\label{appendix-d-evidence-index}

Table 8 indexes every load-bearing claim in the body to the ledger
evidence that verifies it.

\needspace{0.26\textheight}

\textbf{Table 8 \textbar{} Evidence index: each load-bearing claim
mapped to its verifying ledger evidence.}

{\def\LTcaptype{none} 
\begin{longtable}[]{@{}
  >{\raggedright\arraybackslash}p{(\linewidth - 4\tabcolsep) * \real{0.2887}}
  >{\raggedright\arraybackslash}p{(\linewidth - 4\tabcolsep) * \real{0.1856}}
  >{\raggedright\arraybackslash}p{(\linewidth - 4\tabcolsep) * \real{0.5258}}@{}}
\toprule\noalign{}
\begin{minipage}[b]{\linewidth}\raggedright
Claim
\end{minipage} & \begin{minipage}[b]{\linewidth}\raggedright
Evidence
\end{minipage} & \begin{minipage}[b]{\linewidth}\raggedright
Verification
\end{minipage} \\
\midrule\noalign{}
\endhead
\bottomrule\noalign{}
\endlastfoot
Proto-CANONIC origin & \texttt{dividends:07a5834} &
\texttt{git\ show\ 07a5834} \\
First CANON.md & \texttt{canonic:11affab} &
\texttt{git\ show\ 11affab} \\
LANGUAGE.md v0.1 & \texttt{canonic:81bb6d5} &
\texttt{git\ show\ 81bb6d5} \\
10 repositories (window close) & Directory listing & \texttt{ls\ -d\ */}
at the closing ref \\
20 scopes (window close) & CANON.md count &
\texttt{find\ .\ -name\ "CANON.md"} \\
Evidence layer, episodes & Episode count &
\texttt{find\ .\ -name\ "ep*.md"} --- 1 at close, grows at
\texttt{HEAD} \\
Evidence layer, disclosures & Disclosure count &
\texttt{ls\ patents/disclosures/} --- 0 at close, grows at
\texttt{HEAD} \\
\end{longtable}
}

\textbf{Supplemental Material:} the public \texttt{canonic-machine} and
\texttt{canonic-magic} organizations
(https://github.com/canonic-machine) hold the validator source, the
invention-disclosure records, and the complete git history over the
evidence window. Every quantitative claim in this paper resolves to a
command listed in Appendix C run against those repositories; the
supplement is the artifact, not a summary of it.

\subsection{Sources}\label{sources}

Table 9 is the bidirectional citation index: every named entity, numeric
claim, and external artifact in the body resolves below to a stable URL.

\needspace{0.26\textheight}

\textbf{Table 9 \textbar{} Bidirectional citation index: every named
entity and numeric claim mapped to its body anchor and a stable source
URL.}

{\def\LTcaptype{none} 
\begin{longtable}[]{@{}
  >{\raggedright\arraybackslash}p{(\linewidth - 4\tabcolsep) * \real{0.5312}}
  >{\raggedright\arraybackslash}p{(\linewidth - 4\tabcolsep) * \real{0.1042}}
  >{\raggedright\arraybackslash}p{(\linewidth - 4\tabcolsep) * \real{0.3646}}@{}}
\toprule\noalign{}
\begin{minipage}[b]{\linewidth}\raggedright
Claim or named entity
\end{minipage} & \begin{minipage}[b]{\linewidth}\raggedright
Body anchor
\end{minipage} & \begin{minipage}[b]{\linewidth}\raggedright
Source URL
\end{minipage} \\
\midrule\noalign{}
\endhead
\bottomrule\noalign{}
\endlastfoot
Oxford ``AI slop'' Word of the Year 2025 & §Abstract, §1 &
https://languages.oup.com/word-of-the-year/2025/ \\
{[}4{]} Weidinger 2021 --- ethical risks of LMs & §1 &
https://arxiv.org/abs/2112.04359 \\
{[}5{]} Ji 2023 --- hallucination survey & §1 &
https://doi.org/10.1145/3571730 \\
{[}6{]} Bender 2021 --- Stochastic Parrots & §1 &
https://doi.org/10.1145/3442188.3445922 \\
{[}7{]} Backus 1960 --- ALGOL 60 report & §2 &
https://doi.org/10.1145/367236.367262 \\
{[}18{]} Chomsky 1956 --- grammar hierarchy & §2 &
https://doi.org/10.1109/TIT.1956.1056813 \\
{[}14{]} Chomsky 1959 --- formal properties of grammars & §2 &
https://doi.org/10.1016/S0019-9958(59)90362-6 \\
{[}15{]} Knuth 1965 --- LR parsing & §2 &
https://doi.org/10.1016/S0019-9958(65)90426-2 \\
{[}16{]} Aho et al.~2006 --- Dragon Book & §4 &
https://suif.stanford.edu/dragonbook/ \\
{[}19{]} Johnston 1971 --- contour model for block-structured scopes &
§4 & https://doi.org/10.1145/942582.807990 \\
{[}41{]} Dijkstra 1960 --- recursive programming & §4 &
https://doi.org/10.1007/BF01386232 \\
{[}42{]} Landin 1964 --- mechanical expression eval & §4 &
https://doi.org/10.1093/comjnl/6.4.308 \\
{[}20{]} Milner 1978 --- type polymorphism; §4 notes CANONIC shares its
closure property but not its inference & §4, §B.2 &
https://doi.org/10.1016/0022-0000(78)90014-4 \\
{[}40{]} Hindley 1969 --- principal type-scheme & §4 &
https://doi.org/10.1090/S0002-9947-1969-0253905-6 \\
{[}43{]} Cardelli \& Wegner 1985 --- type abstraction & §4 &
https://doi.org/10.1145/6041.6042 \\
{[}44{]} Pierce 2002 --- TAPL & §4 &
https://mitpress.mit.edu/9780262162098/ \\
{[}45{]} Curry \& Feys 1958 --- combinatory logic & §B.2 &
https://archive.org/details/combinatorylogic0001curr \\
{[}47{]} Martin-Löf 1984 --- intuitionistic type theory & §B.2 &
https://archive-pml.github.io/martin-lof/pdfs/Bibliopolis-Book-1984.pdf \\
{[}48{]} Coquand \& Huet 1988 --- calculus of constructions & §B.2 &
https://doi.org/10.1016/0890-5401(88)90005-3 \\
{[}49{]} Lamport 1978 --- distributed clocks & §4 &
https://doi.org/10.1145/359545.359563 \\
{[}50{]} Merkle 1988 --- digital signature & §4 &
https://doi.org/10.1007/3-540-48184-2\_32 \\
{[}51{]} Nakamoto 2008 --- Bitcoin whitepaper & §5 &
https://bitcoin.org/bitcoin.pdf \\
{[}52{]} Torvalds \& Hamano 2005 --- Git & §5 & https://git-scm.com/ \\
{[}21{]} Ritchie 1993 --- development of C & §5 &
https://doi.org/10.1145/154766.155580 \\
{[}22{]} Gosling 2021 --- Java SE 17 specification & §5 &
https://docs.oracle.com/javase/specs/jls/se17/html/index.html \\
{[}23{]} Bradbury 2025 --- Go specification & §5 &
https://go.dev/ref/spec \\
{[}24{]} Klabnik \& Nichols 2023 --- Rust book & §5 &
https://doc.rust-lang.org/book/ \\
{[}2{]} Vaswani 2017 --- transformers & §2 &
https://arxiv.org/abs/1706.03762 \\
{[}3{]} Brown 2020 --- GPT-3 & §2 & https://arxiv.org/abs/2005.14165 \\
Anthropic Claude Opus model family (4.5 production, 4.7 byline
reconciliation) & §Disclosure, {[}53{]} &
https://www.anthropic.com/claude \\
{[}54{]} OpenAI GPT-4 technical report & §2 &
https://arxiv.org/abs/2303.08774 \\
{[}55{]} Ostrom 1990 --- governing the commons & §5 &
https://doi.org/10.1017/CBO9780511807763 \\
{[}56{]} Lessig 1999 --- Code and Other Laws & §5 &
https://archive.org/details/codeotherlawsofc00less \\
{[}57{]} Wright \& De Filippi 2015 --- Lex Cryptographia & §5 &
https://papers.ssrn.com/sol3/papers.cfm?abstract\_id=2580664 \\
{[}29{]} Hadley 2017 --- drug repurposing (eLife) & §Author Context &
https://pubmed.ncbi.nlm.nih.gov/28936969/ \\
{[}30{]} Hadley 2017 --- STARGEO precision annotation of digital samples
& §Author Context & https://pubmed.ncbi.nlm.nih.gov/28925997/ \\
{[}31{]} Panahiazar 2022 --- MammoChat AI & §Author Context &
https://pubmed.ncbi.nlm.nih.gov/34697751/ \\
{[}32{]} Ding 2019 --- large-scale labeling & §Author Context &
https://pubmed.ncbi.nlm.nih.gov/30128778/ \\
{[}34{]} Wong 2018 --- delirium-risk ML & §Author Context &
https://pubmed.ncbi.nlm.nih.gov/30646095/ \\
{[}35{]} Ding 2019 --- Alzheimer's PET deep learning & §Author Context &
https://pubmed.ncbi.nlm.nih.gov/30398430/ \\
{[}36{]} Wang 2007 --- PennCNV & §Author Context &
https://pubmed.ncbi.nlm.nih.gov/17921354/ \\
{[}39{]} MammoChat --- empathic, patient-centered, blockchain provenance
& §5, §5 & https://github.com/canonic-machine/mammochat \\
{[}63{]} STARGEO repo --- Search Tag Analyze Resource for GEO & §Author
Context & https://github.com/idrdex/stargeo \\
{[}64{]} HadleyLab repos --- Translating Big Data into Precision
Medicine & §Author Context & https://github.com/hadleylab \\
{[}58{]} CANONIC LANGUAGE spec --- defines lexical grammar after Aho,
with block-structured scopes & §5 &
https://github.com/canonic-machine/canonic \\
{[}61{]} CANONIC VALIDATORS --- open-source predicate binaries for
Apache-licensed enforcement & §3 &
https://github.com/canonic-machine/VALIDATORS \\
{[}65{]} CANONIC Foundation --- stewards the AI-First Human-Governed
paradigm & §5 & https://canonic.org/ \\
NIH U01-LM012675 (CrADLe) & §Author Context &
https://reporter.nih.gov/search/?queryText=U01LM012675 \\
NIH BD2K program (2013--2020) & §Author Context &
https://commonfund.nih.gov/bd2k \\
ORCID 0000-0003-0990-4674 (Hadley) & §Byline, §Disclosure &
https://orcid.org/0000-0003-0990-4674 \\
\end{longtable}
}

\end{document}